\documentclass[preprint,5pt]{elsarticle}
\usepackage[T1]{fontenc}
\usepackage{amsmath}
\usepackage{graphicx}
\usepackage{epstopdf}
\usepackage{subcaption}
\usepackage{amssymb}
\usepackage{esint}
\usepackage{amsthm}
\usepackage{dcolumn}
\usepackage{bm}
\usepackage{multirow}
\usepackage{geometry}
\usepackage{booktabs}
\usepackage{longtable}
\journal{Laser Physics Letters}
\begin{document}
	\newtheorem{conjecture}{Conjecture}\newtheorem{corollary}{Corollary}\newtheorem{theorem}{Theorem}
	\newtheorem{lemma}{Lemma}\newtheorem{observation}{Observation}\newtheorem{definition}{Definition}
	\newtheorem{remark}{Remark}\global\long\global\long\def\ket#1{|#1 \rangle}
	\global\long\global\long\def\bra#1{\langle#1|}
	\global\long\global\long\def\proj#1{\ket{#1}\bra{#1}}
	\begin{frontmatter}

		\title{Degenerate perturbation theory to quantum search}

		\author[gc1]{Dezheng Zhang}
		\author[gc1]{Xuanmin Zhu\corref{cor}}
		\ead{zhuxuanmin2006@163.com}
		\author[gc1]{Yuanchun Deng}
		\author[gc1]{Runping Gao}
		\author[xd1]{Qun wei}
		\author[gc2]{Zijiang Luo}
		\address[gc1]{Center for Quantum Information, School of Information, Guizhou University of Finance and Economics, Guiyang, Guizhou 550025, China}

		\address[xd1]{School of Physics and Optoelectronic Engineering, Xidian University, Xi'an 710071, China}
		\address[gc2]{Institute of Intelligent Manufacturing, Shunde Polytechnic, Guangdong Shunde 528300,China}
		\cortext[cor]{Corresponding author. Tel.: +86 15802914790.}
		\begin{abstract}
We utilize degenerate perturbation theory to investigate continuous-time quantum search on second-order truncated simplex lattices. In this work, we show that the construction of the Hamiltonian must consider the structure of the lattice. This idea enables effective application of degenerate perturbation theory to third- and higher-order lattices. We identify two constraints on the reduction of the dimension of the Hamiltonian. In addition, we elucidate the influence of the distinct configurations of marked vertices on the quantum search.
		\end{abstract}
		
		\begin{keyword}
	Grover's algorithm  \sep Quantum search \sep  Quantum random walk \sep Continuous-time quantum walk \sep Truncated simplex lattice
			
		\end{keyword}
		
	\end{frontmatter}
	
	\section{Introduction}	
	Grover's algorithm is a powerful algorithm utilized for quantum search in an unstructured database~\cite{groveralgorithm,searchforaneedle,analoganalogue}. Continuous-time quantum walk (CTQW) was introduced to address the search problems in the structured databases by quantizing the classical walk through simulating the classical Markov process~\cite{ctqwanddecisiontrees}. Continuous-time quantum walk has successfully solved the quantum search problems on complete graphs, hypercubes, d-dimensional lattice graphs and other types of graphs with substantial speedup~\cite{childs2004,globalsymmetry,connectivity,almostallgraphs,johnsongraphs,quadraticspeedup}.
	Moreover, it has been extended to information transmission in networks~\cite{almostallgraphs,informationpropagation}.
	
	Truncated simplex lattices are of particular interest due to their effectively nonintegral dimensionality. Continuous-time quantum walks have also gained considerable attention for their application in quantum search algorithms on truncated simplex lattices~\cite{lattices,minimization,statistics}. Initially, the study of the quantum search in this area was focused on zeroth-order lattices~\cite{childs2004}.  Subsequently, Thomas G Wong studied first-order truncated simplex lattices~\cite{connectivity}, and Wang Yunkai et al. studied quantum search on second and higher-order lattices~\cite{optimalquantumsearch,roleofsymmetry}. Zhu Xuanmin et al. explored quantum search for searching a set of marked vertices~\cite{sisters}.
	
	In the CTQW, the evolution of the system relies on the jumping rate $\gamma$ which represents the probability of the transition between the adjacent vertices per unit time~\cite{fromquantumcellular,ontheabsence}. Taking an appropriate value at $\gamma$, known as the critical jumping rate $\gamma_c$, the system can evolve into the target state at an appropriate time.
	The common approach for determining the critical jumping rate $\gamma_c$ involves calculating the squared overlaps of the Hamiltonian's eigenstates with the basis states~\cite{childs2004,connectivity,optimalquantumsearch,roleofsymmetry,sisters}. As the order of the lattice increases, this method can not be easily implemented.
	
	In order to determine $\gamma_{c}$ more accurately, Thomas G. Wong introduced degenerate perturbation theory to continuous-time quantum walk~\cite{jjmodernquantum,degenerateperturbationtheory,diagrammaticapproach}.
	He provided the schemes of degenerate perturbation theory to determine $\gamma_{c}$ on complete graphs, first-order truncated simplex lattices, and hypercubes~\cite{latticesof}.
	In cases involving weighted graphs and multiple configurations of marked vertices, quantum search schemes utilizing degenerate perturbation theory have also been proposed~\cite{weightedgraphs,multiplemarkers}.
	
	However, the existing schemes can not be used directly to obtain the $\gamma_{c}$ on the second-order and higher-order lattices. To address this, we present an scheme for implementing degenerate perturbation theory in quantum searches on second-order and third-order lattices.
	The results of our investigations emphasize the importance of the lattice structure when constructing the leading-order terms of Hamiltonian. The dimension of the Hamiltonian can also be reduced to simplify the computational process by eliminating uncorrelated vertices in the evolution. The influence of different configurations of marked vertices on the quantum search is further explored by six simulation experiments with distinct configurations on the second-order lattice.
	
	This paper is structured as follows. In Section \ref{sec:secondorder}, we explore the application of degenerate perturbation theory to quantum search on a second-order lattice. We study the quantum search on third-order lattices, and present two constraints aimed at reducing the dimension of the Hamiltonian in Section \ref{sec:rules}. In Section \ref{sec:theinfluence}, we discuss quantum search scenarios with different configurations of marked vertices on the second-order lattice. Finally, we give our conclusions in Section \ref{sec:conclusion}.
\begin{figure}[htbp]
	\centering
	\begin{subfigure}{0.4\linewidth}
		\centering
		\includegraphics[width=0.8\linewidth]{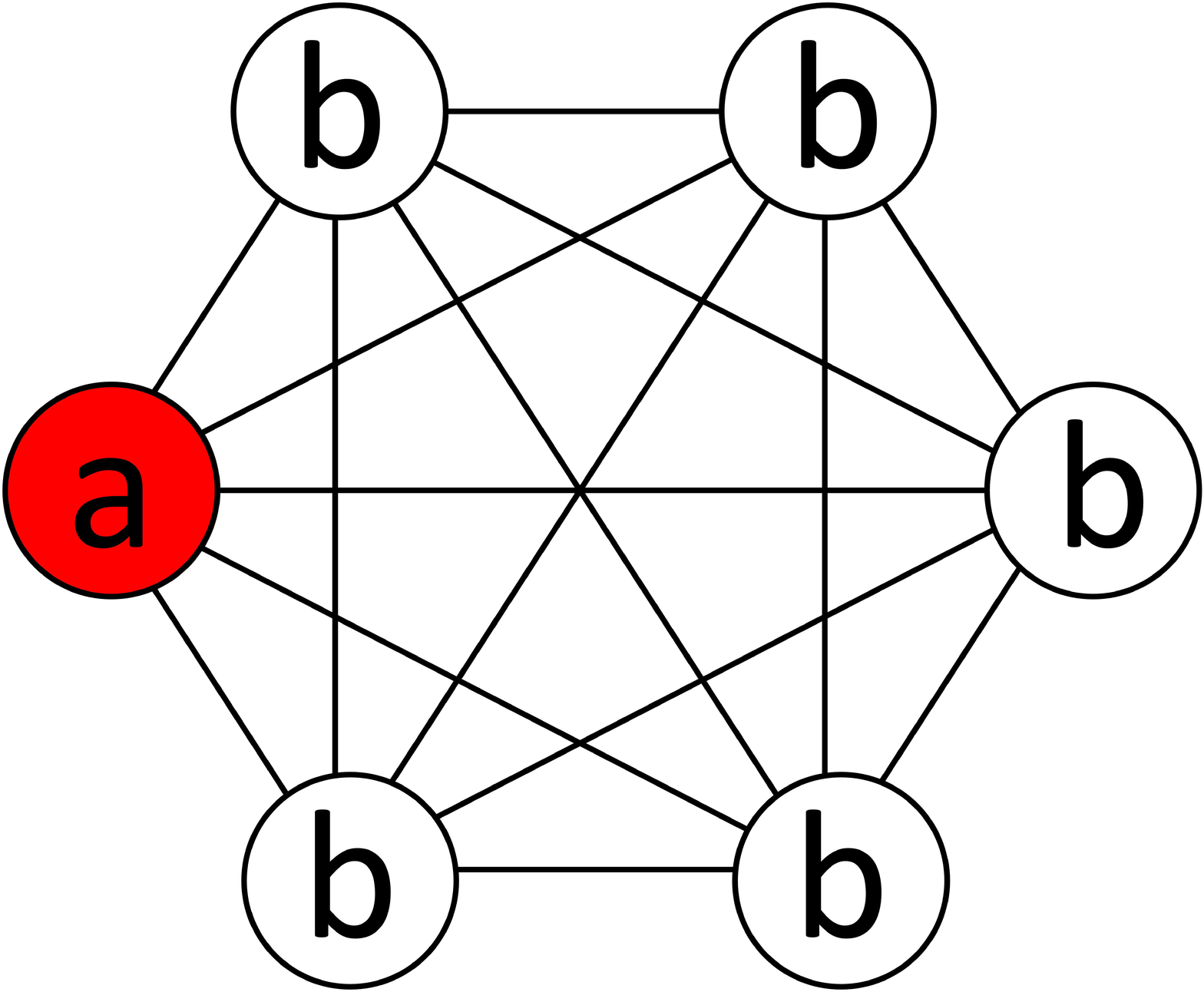}
		\caption{}
		\label{completegraph}
	\end{subfigure}
	\begin{subfigure}{0.4\linewidth}
		\centering
		\includegraphics[width=0.7\linewidth]{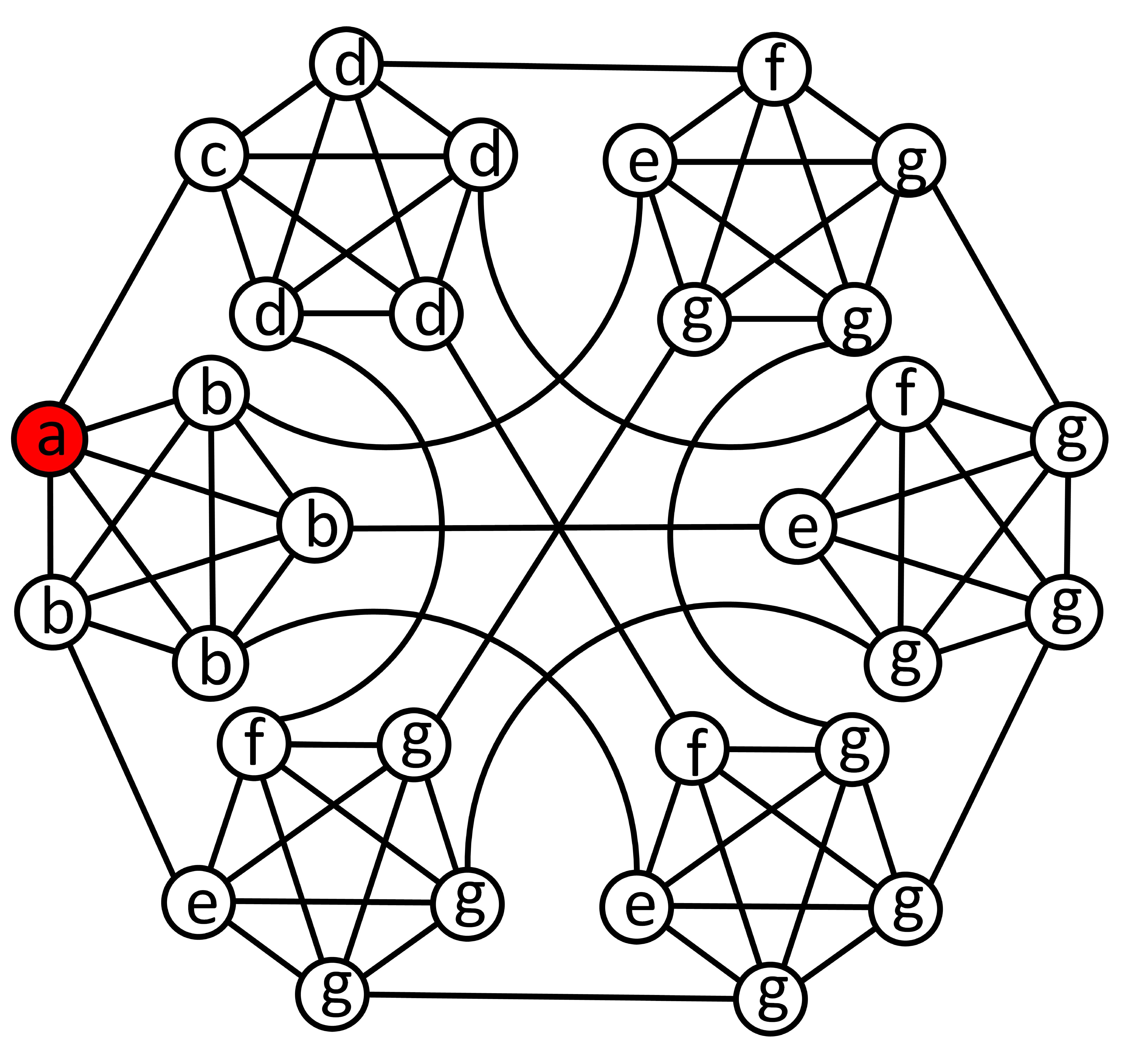}
		\caption{}
		\label{firstorder}
	\end{subfigure}
	\caption{\textbf{a} A zeroth-order $M$-dimensional simplex lattice. \textbf{b} A first-order $M$-dimensional simplex lattice. $M=5$.}
	\label{graphH0}
\end{figure}
	\begin{figure}[htbp]
		\centering
		\includegraphics[scale=0.04]{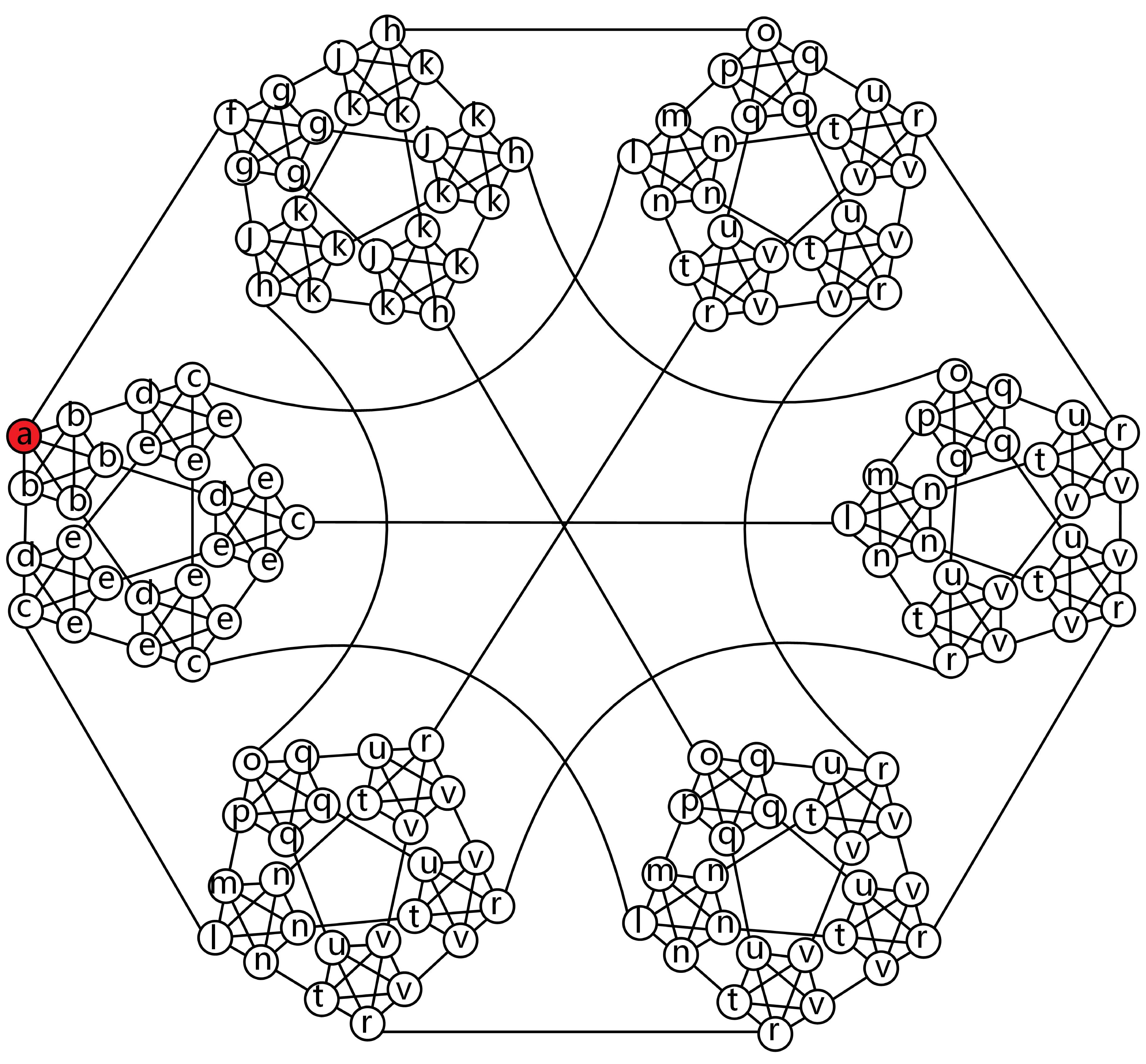}
		\caption{A second-order truncated five-dimensional simplex lattice. Vertices that evolving identically are labelled with the same letter. The red vertex labeled $a$ corresponds to the marked state $\left| a \right\rangle$.} \label{second-order}
\end{figure}
	
\section{Quantum search on the second-order lattices}
	\label{sec:secondorder}

	Truncated simplex lattices are derived from complete graphs. A complete graph consisting of $M$ vertices can be referred to as a $M$-dimensional complete graph. And a $(M+1)$-dimensional complete graph can be referred to as a zeroth-order truncated $M$-dimensional simplex lattice, as illustrated in Fig.~\ref{completegraph}.
	A first-order $M$-dimensional lattice is obtained by replacing each vertex in the zeroth-order lattice with a $M$-dimensional complete graph, resulting in a total of $N = M(M+1)$ vertices, as shown in Fig. \ref{firstorder}.
	Similarly, a second-order $M$-dimensional lattice is obtained by replacing each vertex in a first-order lattice with a $M$-dimensional complete graph, resulting in a total of $N = M^2(M+1)$ vertices, as shown in Fig. \ref{second-order}~\cite{lattices}.
	
	 Without the marked vertex, the Hamiltonian of the system can be expressed as follows:
	\begin{equation}
		H=-\gamma L. \label{gammaL}
	\end{equation}
	$L = A - D$ is the graph Laplacian~\cite{multiplemarkers}, where $A$ is the adjacency matrix ($A_{ij} = 1$ if the vertex $i$ is connected to vertex $j$ directly, and $A_{ij}=0$ otherwise), and $D$ is the diagonal degree matrix ($D_{ij}=deg(i)$). With the marked vertex, the Hamiltonian can be represented as~\cite{oracles}:
	\begin{equation}
		H=-\gamma L- \sum_{i \in marked} \left| i \right\rangle\left\langle i \right|,
	\end{equation}
	where $|i\rangle$ is the marked state.	
	\begin{table}[htbp]
		\renewcommand\arraystretch{0.5}
		\centering
		\caption{The adjacency matrix $A$ of a second-order $M$-dimensional simplex lattice in the invariant subspace, where ${M_l} = M - l$ .}\label{second-adjacency}
		\setlength{\tabcolsep}{0mm}{
			\small
			\begin{longtable}{cccccccccccccccccccc}
				\hline
				\hline
				\\
				0&$\sqrt {{M_1}}$&0&0&0&1&0&0&0&0&0&0&0&0&0&0&0&0&0&0\\
				$\sqrt {{M_1}}$&${M_2}$&0&1&0&0&0&0&0&0&0&0&0&0&0&0&0&0&0&0\\
				0&0&0&1&$\sqrt {{M_2}}$&0&0&0&0&0&1&0&0&0&0&0&0&0&0&0\\
				0&1&1&0&$\sqrt {{M_2}}$&0&0&0&0&0&0&0&0&0&0&0&0&0&0&0\\
				0&0&$\sqrt {{M_2}}$&$\sqrt {{M_2}}$&${M_2}$&0&0&0&0&0&0&0&0&0&0&0&0&0&0&0\\
				1&0&0&0&0&0&$\sqrt {{M_1}}$&0&0&0&0&0&0&0&0&0&0&0&0&0\\
				0&0&0&0&0&$\sqrt {{M_1}}$&${M_2}$&0&1&0&0&0&0&0&0&0&0&0&0&0\\
				0&0&0&0&0&0&0&0&1&$\sqrt {{M_2}}$&0&0&0&1&0&0&0&0&0&0\\
				0&0&0&0&0&0&1&1&0&$\sqrt {{M_2}}$&0&0&0&0&0&0&0&0&0&0\\
				0&0&0&0&0&0&0&$\sqrt {{M_2}}$&$\sqrt {{M_2}}$&${M_2}$&0&0&0&0&0&0&0&0&0&0\\
				0&0&1&0&0&0&0&0&0&0&0&1&$\sqrt {{M_2}}$&0&0&0&0&0&0&0\\
				0&0&0&0&0&0&0&0&0&0&1&0&$\sqrt {{M_2}}$&0&1&0&0&0&0&0\\
				0&0&0&0&0&0&0&0&0&0&$\sqrt {{M_2}}$&$\sqrt {{M_2}}$&${M_3}$&0&0&0&0&1&0&0\\
				0&0&0&0&0&0&0&1&0&0&0&0&0&0&1&$\sqrt {{M_2}}$&0&0&0&0\\
				0&0&0&0&0&0&0&0&0&0&0&1&0&1&0&$\sqrt {{M_2}}$&0&0&0&0\\
				0&0&0&0&0&0&0&0&0&0&0&0&0&$\sqrt {{M_2}}$&$\sqrt {{M_2}}$&${M_3}$&0&0&1&0\\
				0&0&0&0&0&0&0&0&0&0&0&0&0&0&0&0&1&1&1&$\sqrt {{M_3}}$\\
				0&0&0&0&0&0&0&0&0&0&0&0&1&0&0&0&1&0&1&$\sqrt {{M_3}}$\\
				0&0&0&0&0&0&0&0&0&0&0&0&0&0&0&1&1&1&0&$\sqrt {{M_3}}$\\
				0&0&0&0&0&0&0&0&0&0&0&0&0&0&0&0&$\sqrt {{M_3}}$&$\sqrt {{M_3}}$&$\sqrt {{M_3}}$&${M_3}$\\
				\\
				\hline
				\hline
			\end{longtable}
		}
	\end{table}
	
	\newpage
	In regular graphs, all vertices have the same degree. The diagonal matrix $D$ becomes a multiple of the identity matrix, and it does not influence the quantum search. Then the Laplacian operator $L$ can be replaced by the adjacency matrix $A$~\cite{laplacianvsadjacency}. The Hamiltonian can be expressed as $H=-\gamma A- \sum_{i \in marked} \left| i \right\rangle\left\langle i \right|$.	
	
	The initial state is chosen the one with the equal probability distributed among the $N$ vertices, as: $\left| {{\psi}}(0) \right\rangle = \frac{1}{\sqrt{N}} \sum_{i=1}^{N} \left| i \right\rangle$. While $\left| a \right\rangle$ is the marked state, based on the symmetry of the lattice, the second-order lattice can be seen as a system evolving in a $20$-dimensional invariant subspace~\cite{roleofsymmetry}. This subspace consists of the following basis states: $\left| a \right\rangle$, $\left| b \right\rangle = \frac{1}{\sqrt{M-1}} \sum_{i \in b} \left| i \right\rangle$, $\left| e \right\rangle = \frac{1}{\sqrt{(M-2)(M-1)}} \sum_{i \in e} \left| i \right\rangle$, $\left| v \right\rangle = \frac{1}{\sqrt{(M-3)(M-2)(M-1)}} \sum_{i \in v} \left| i \right\rangle$, and so on.
	The vertices corresponding to different states are represented by $20$ different letters, and vertices with the same evolution are denoted by the same letter, as shown in Fig. \ref{second-order}.
The Hamiltonian is
		\begin{equation}
			H = -\gamma A - \left| a \right\rangle\left\langle a \right|
		\end{equation}
	where $\left| a \right\rangle\left\langle a \right|$ is considered as an quantum oracle. The adjacency matrix expressed in the 20-dimensional subspace is shown in Table~\ref{second-adjacency}.

	\begin{figure}[htbp]
		\centering
		\begin{subfigure}{0.45\linewidth}
			\centering
			\includegraphics[width=0.9\linewidth]{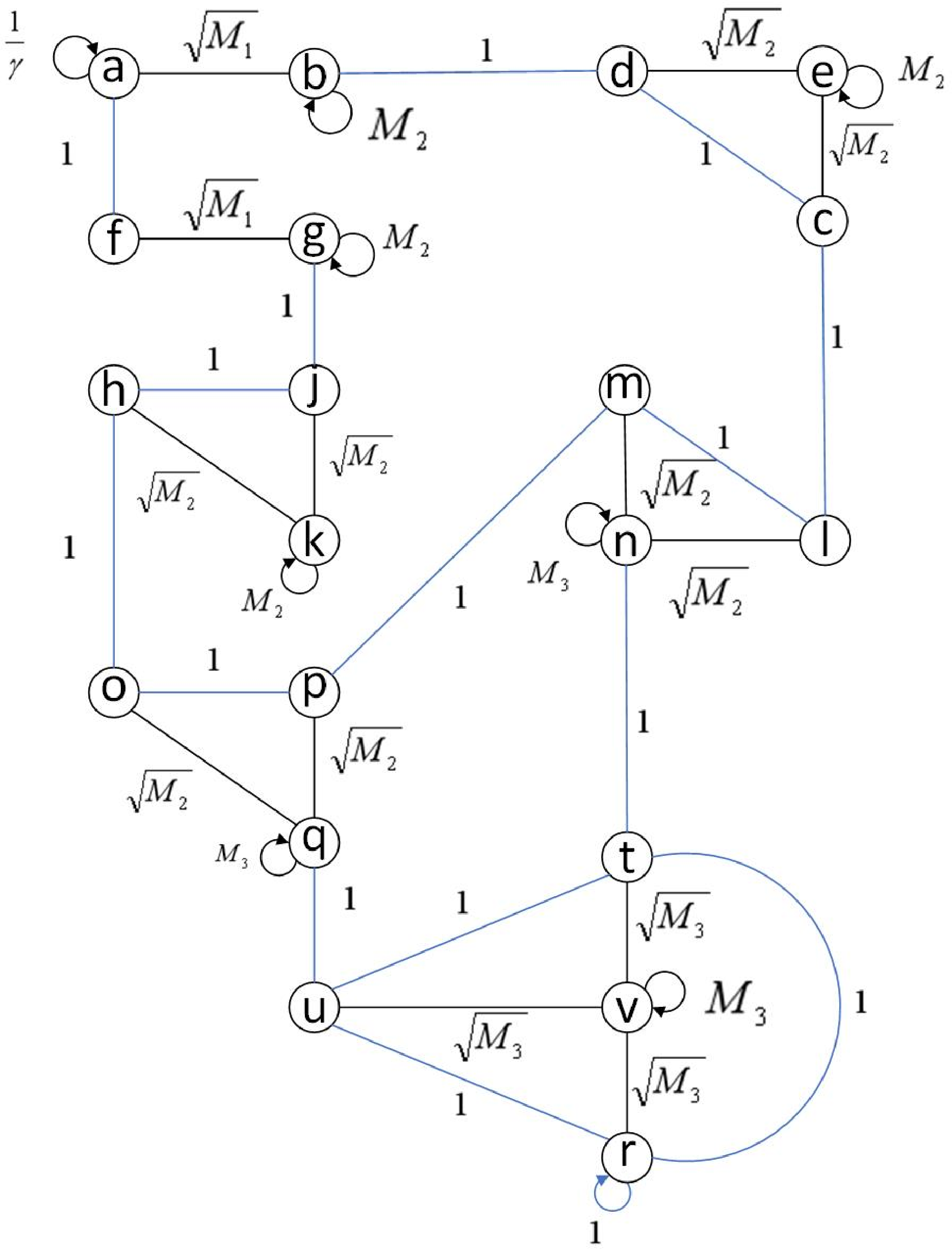}
			\caption{}
			\label{Hamiltonian}
		\end{subfigure}
		\centering
		\begin{subfigure}{0.45\linewidth}
			\centering
			\includegraphics[width=0.9\linewidth]{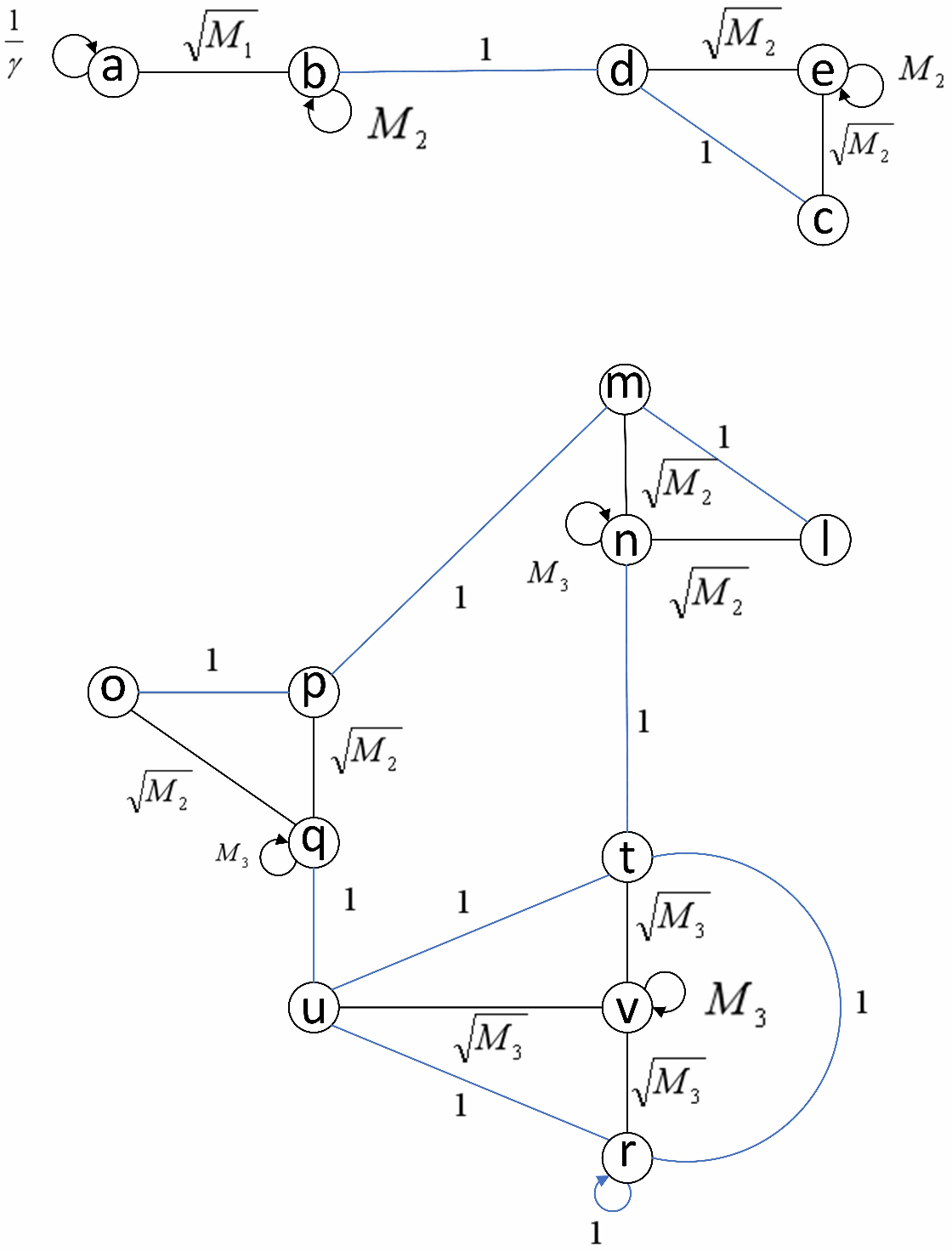}
			\caption{}
			\label{H0_1}
		\end{subfigure}
		\centering
		\begin{subfigure}{0.45\linewidth}
			\centering
			\includegraphics[width=0.9\linewidth]{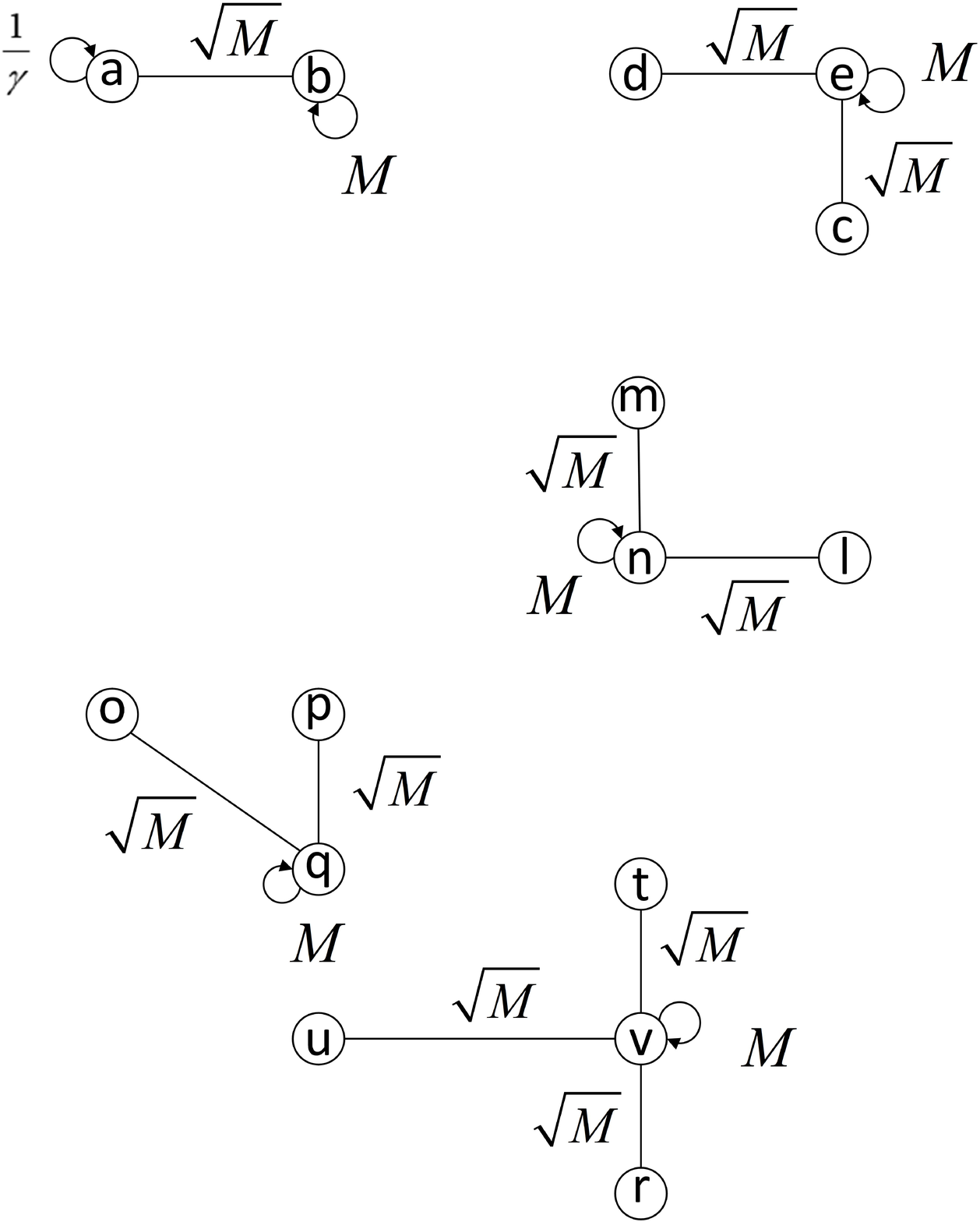}
			\caption{}
			\label{H0_2}
		\end{subfigure}
		\centering
		\begin{subfigure}{0.45\linewidth}
			\centering
			\includegraphics[width=0.9\linewidth]{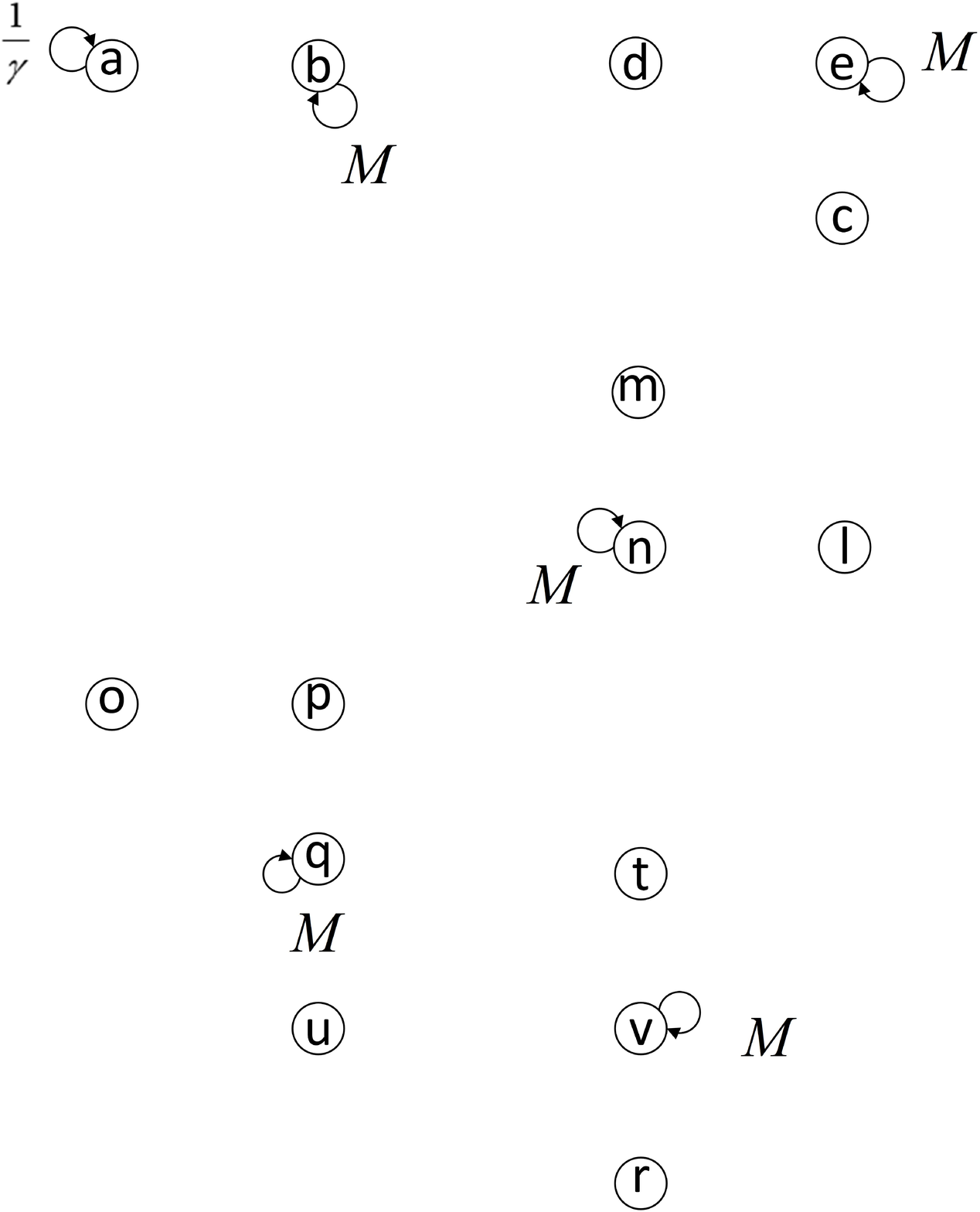}
			\caption{}
			\label{H0_3}
		\end{subfigure}
		\caption{\textbf{a} A graphical representation of the Hamiltonian for the second-order truncated five-dimensional simplex lattice. \textbf{b} The leading-order term for the first stage of the algorithm. \textbf{c} The leading-order term for the second stage of the algorithm. \textbf{d} The leading-order term for the third stage of the algorithm.}
		\label{graphH0}
	\end{figure}
	
	Since $\left| {{\psi}}(0) \right\rangle  \approx \left| v \right\rangle$ for large $M$, the objective of the quantum search is to achieve the system evolution from $\left| v \right\rangle$ to $\left| a \right\rangle$. The quantum search on the second-order lattice is structured into three sequential stages: from $\left| v \right\rangle$ to $\left| e \right\rangle$, then to $\left| b \right\rangle$, and ultimately to $\left| a \right\rangle$. We denote the critical jumping rates of these stages as $\gamma_{c1}$, $\gamma_{c2}$, and $\gamma_{c3}$, respectively.
	
\begin{figure}
		\centering
		\includegraphics[scale=0.5]{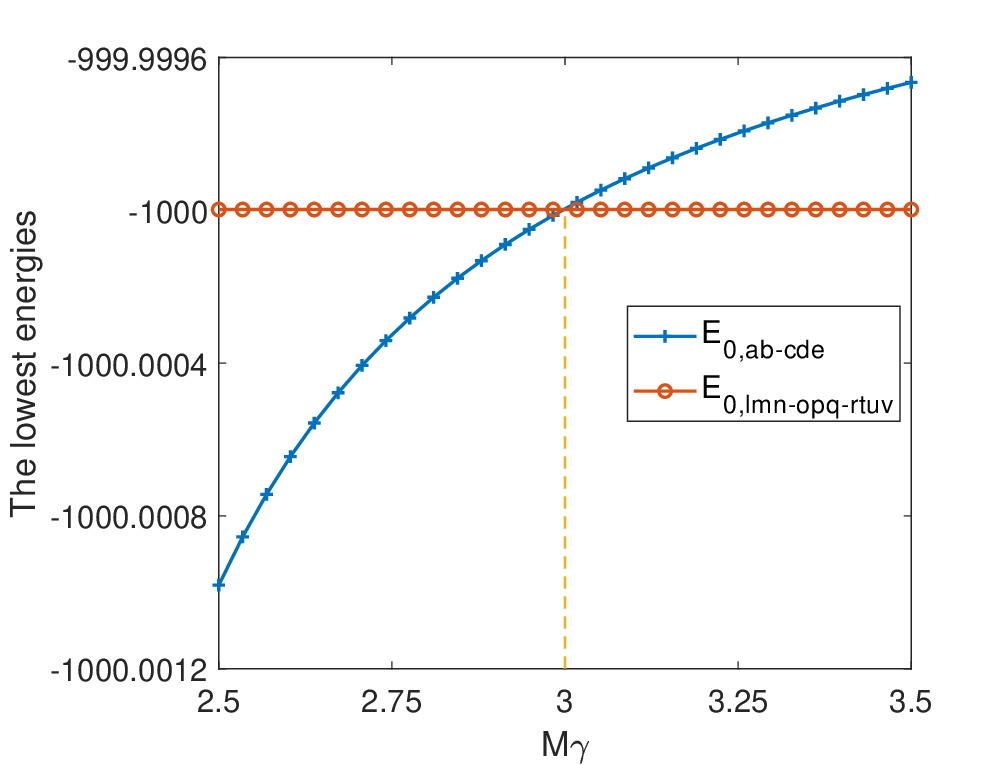}
		\caption{$\gamma_{c1}$  is determined when  $E_{0,ab-cde} = E_{0,lmn-opq-rtuv}$.}\label{gamma_1ofsecondorder}
	\end{figure}

	The search Hamiltonian can be represented by a weighted graph, as shown in Fig. \ref{Hamiltonian}~\cite{diagrammaticapproach}. Based on the lattice structure, the sets $\{ \left| f \right\rangle,  \left| g \right\rangle,  \left| h \right\rangle,  \left| j \right\rangle,  \left| k \right\rangle\}$ are deemed irrelevant to the search. Therefore, we disregard their presence in the weighted graph and focus solely on the last part depicted in Figure \ref{H0_1}, which serves as the leading-order term $H_1^{(0)}$ in the first stage. The leading-order term $H_1^{(0)}$ can be expressed as
	\begin{equation}
		H_1^{(0)} = \left(
		\begin{array}{cc}
			H_{ab-cde}^{(0)} & 0\\
			0 & H_{lmn-opq-rtuv}^{(0)}\\
		\end{array} \right),\label{H0ofstage1}
	\end{equation}
	where
	\begin{equation}
		H_{ab-cde}^{(0)} = -\gamma \left(
		\begin{array}{ccccc}
			\frac{1}{\gamma} & \sqrt{M_1} & 0 & 0 & 0 \\
			\sqrt{M_1} & M_2 & 0 & 1 & 0 \\
			0 & 0 & 0 & 1 & \sqrt{M_2} \\
			0 & 1 & 1 & 0 & \sqrt{M_2} \\
			0 & 0 & \sqrt{M_2} & \sqrt{M_2} & M_2\\
		\end{array} \right),
	\end{equation}
	\begin{equation}
		H_{lmn-opq-rtuv}^{(0)} = -\gamma \left(
		\begin{array}{cccccccccc}
			0 & 1 & \sqrt{M_2} & 0 & 0 & 0 & 0 & 0 & 0 & 0 \\
			1 & 0 & \sqrt{M_2} & 0 & 1 & 0 & 0 & 0 & 0 & 0 \\
			\sqrt{M_2} & \sqrt{M_2} & M_3 & 0 & 0 & 0 & 0 & 1 & 0 & 0 \\
			0 & 0 & 0 & 0 & 1 & \sqrt{M_2} & 0 & 0 & 0 & 0 \\
			0 & 1 & 0 & 1 & 0 & \sqrt{M_2} & 0 & 0 & 0 & 0 \\
			0 & 0 & 0 & \sqrt{M_2} & \sqrt{M_2} & M_3 & 0 & 0 & 1 & 0 \\
			0 & 0 & 0 & 0 & 0 & 0 & 1 & 1 & 1 & \sqrt{M_3} \\
			0 & 0 & 0 & 0 & 0 & 0 & 1 & 0 & 1 & \sqrt{M_3} \\
			0 & 0 & 0 & 0 & 0 & 1 & 1 & 1 & 0 & \sqrt{M_3} \\
			0 & 0 & 0 & 0 & 0 & 0 & \sqrt{M_3} & \sqrt{M_3} & \sqrt{M_3} & M_3 \\
		\end{array} \right).
	\end{equation}
	The eigenstates of $H_{ab-cde}^{(0)}$ and $H_{lmn-opq-rtuv}^{(0)}$ can be expressed as
	\begin{equation}
		\left| sp_{ab-cde} \right\rangle=\alpha_a \left| a \right\rangle + \alpha_b \left| b \right\rangle + \alpha_c \left| c \right\rangle + \alpha_d \left| d \right\rangle + \alpha_e \left| e \right\rangle ,
	\end{equation}
	\begin{equation}
		\left| sp_{lmn-opq-rtuv} \right\rangle=\alpha_l \left| l \right\rangle + \alpha_m \left| m \right\rangle + \alpha_n \left| n \right\rangle + \alpha_o \left| o \right\rangle + \alpha_p \left| p \right\rangle + \alpha_q \left| q \right\rangle + \alpha_r \left| r \right\rangle + \alpha_t \left| t \right\rangle + \alpha_u \left| u \right\rangle + \alpha_v \left| v \right\rangle .
	\end{equation}
	The eigenvalues $E_{ab-cde}$ and $E_{lmn-opq-rtuv}$ of $H_{ab-cde}^{(0)}$ and $H_{lmn-opq-rtuv}^{(0)}$ are the corresponding energies. To ensure the system evolves from $\left| v \right\rangle$  to $\left| e \right\rangle$, the two lowest energies $E_{0,ab-cde}$ and $E_{0,lmn-opq-rtuv}$ should be equal to each other, and the eigenstates of them are denoted as $\left| spe1 \right\rangle$ and $\left| spv1 \right\rangle$, respectively.
	The two lowest energies are given in Fig. \ref{gamma_1ofsecondorder}, and the critical jumping rate $\gamma_{c1} = 3 / M$. For the second-order lattice, the edges that scale less than $\sqrt{M}$ in Fig. \ref{H0_1} can not be excluded and the approximation $M-l \approx M$ and $\sqrt{M-l} \approx \sqrt{M}$ are not applicable, which are different with the quantum search on the first-order lattices~\cite{diagrammaticapproach}.

	The total Hamiltonian is defined as $H_1 = H_1^{(0)} + H_1^{(1)}$, where the perturbation term $H_1^{(1)}$ includes the edge connecting the vertices $e$ and $l$, symbolizing the connection between the sets $\{a, b, c, d, e\}$ and $\{l, m, n, o, p, q, r, t, u, v\}$. By using the basis $\{\left| spe1 \right\rangle, \left| spv1 \right\rangle\}$, the whole Hamiltonian can be expressed as \begin{small}
		$H_1\!=\! H_{1,spv1,spv1}\!\left| spv1 \rangle\!\langle spv1 \right|\!+\!H_{1,spv1,spe1}\!\left| spv1 \rangle\!\langle spe1 \right|\!+\!H_{1,spe1,spv1}\!\left| spe1 \rangle\!\langle spv1 \right|\!+\!H_{1,spe1,spe1}\!\left| spe1 \rangle\!\langle spe1 \right|.$
	\end{small}
	The eigenstates of $H_1$ on this two-dimensional subspace are
	\begin{align}
		\left| \phi_{1-} \right\rangle & = \frac{1}{\sqrt{2}} (\left| spv1 \right\rangle + \left| spe1 \right\rangle),  \\
		\left| \phi_{1+} \right\rangle & = \frac{1}{\sqrt{2}} (\left| spv1 \right\rangle - \left| spe1 \right\rangle).
	\end{align}

	The eigenvalues are denoted as $E_{1-}$ and $E_{1+}$, respectively. When $M$ is large, $\left| spe1 \right\rangle \approx \left| e \right\rangle$ and $\left| spv1 \right\rangle \approx \left| v \right\rangle$. The system evolves from $\left| v \right\rangle$ to $\left| e \right\rangle$ at the time $t_1 = \pi/(E_{1+} - E_{1-})$.
		
	We represent the success probability of the system evolution from $\left| v \right\rangle$ to $\left| e \right\rangle$ as $P_1$. To study the stability of this algorithm, we assume that the jumping rate $\gamma_{1}$ is $3/M+\epsilon_1$ instead of $3/M$. When $M=100$, the range of $\epsilon_1$ that ensures $P_1 \geqslant 50\%$ is $[ -2 \times 10^{-3}, 2 \times 10^{-3} ]$. Similarly, for $M=1000$ and $M=10000$, in order to ensure $P_1 \geqslant 50\%$, the value of $\epsilon_1$ should fall in the intervals of $[-6 \times 10^{-5}, 6 \times 10^{-5}]$ and $[-2 \times 10^{-6}, 2 \times 10^{-6}]$ respectively. The detailed results are given in Table \ref{gamma1Table}.

	\begin{table}
		\setcounter{table}{1}
		\centering
		\caption{On the second-order lattice, when the target state is $\left| a \right\rangle$, the values of $\gamma_{1}$ are required in order to ensure $P_1 \geqslant 50\%$.}\label{gamma1Table}
		\begin{tabular}{cc p{15em}<{\centering} p{10em}<{\centering}}
			\hline
			$M$ & $\gamma_{c1}=3/M$ & $\gamma_{1}=\gamma_{c1} + \epsilon_1$ & $\frac{|\epsilon_1|}{\gamma_{c1}} \times 100\%$ \\
			\midrule[1pt]
			$100$ & $0.03$ & $2.8 \times 10^{-2} \sim 3.2 \times 10^{-2}$ & $6.67\%$ \\
			$1000$ & $0.003$ & $2.94 \times 10^{-3} \sim 3.06 \times 10^{-3}$ & $2\%$ \\
			$10000$ & $0.0003$ & $2.98 \times 10^{-4} \sim 3.02 \times 10^{-4}$ & $0.67\%$ \\
			\hline
		\end{tabular}
	\end{table}
	
	In the second stage of the algorithm, the system evolves from $\left| \psi(t_1) \right\rangle \approx \left| e \right\rangle$ to $ \left| b \right\rangle$. The leading-order term $H_{2}^{(0)}$ of the Hamiltonian is shown in Fig. \ref{H0_2}.
	We approximate $M-l \approx M$ and $\sqrt{M-l} \approx \sqrt{M}$ while excluding edges with weights less than $\sqrt{M}$.
	The leading-order term of the Hamiltonian can be expressed as
	\begin{equation}
		H_2^{(0)} = \left(
		\begin{array}{cc}
			H_{ab}^{(0)} & 0\\
			0 & H_{cde}^{(0)}\\
		\end{array} \right),
	\end{equation}
	where
	\begin{align}
		H_{ab}^{(0)} &= -\gamma \left(
		\begin{array}{cc}
			1/\gamma & \sqrt{M} \\
			\sqrt{M} & M \\
		\end{array} \right), \\
		H_{cde}^{(0)} &= -\gamma \left(
		\begin{array}{ccc}
			0 & 0 & \sqrt{M} \\
			0 & 0 & \sqrt{M} \\
			\sqrt{M} & \sqrt{M} & M \\
		\end{array} \right).
	\end{align}
		
\begin{figure}
		\centering
		\includegraphics[scale=0.5]{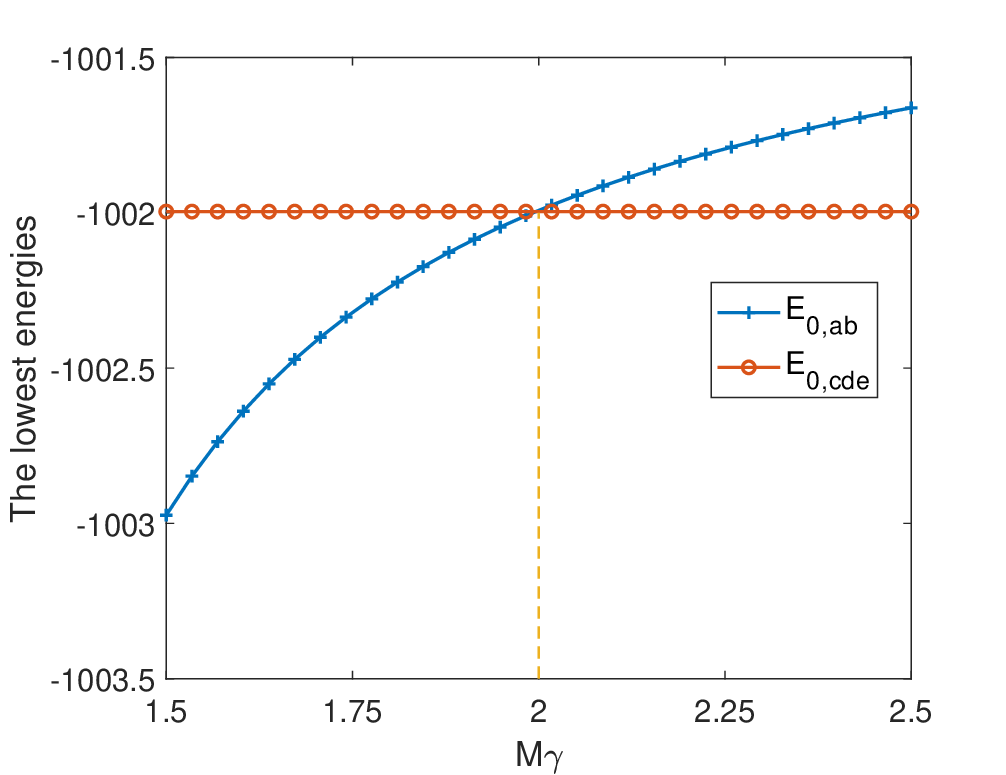}
		\caption{$\gamma_{c2}$ is determined when  $E_{0,ab} = E_{0,cde}$.}\label{gamma_2ofsecondorder}
	\end{figure}

	We obtained $\gamma_{c2}=2/M$ by equating the lowest energy $E_{0,ab}$ of $H_{ab}^{(0)}$ to the lowest energy $E_{0,cde}$ of $H_{cde}^{(0)}$, as depicted in Fig. \ref{gamma_2ofsecondorder}.
	
	By using the whole Hamiltonian $H_{2}=H_{2}^{(0)}+H_2^{(1)}$, we have the eigenstates of the system $\left| \phi_{2\mp} \right\rangle = \frac{1}{\sqrt{2}} (\left| spb2 \right\rangle \pm \left| spe2 \right\rangle)$, where $\left| spb2 \right\rangle$ and $\left| spe2 \right\rangle$ are the eigenvectors of the lowest energies $E_{0,ab}$ and $E_{0,cde}$. And the eigenvalues of $\left| \phi_{2\mp} \right\rangle$ are $E_{2-}$ and $E_{2+}$, respectively. The system evolves from $\left| spe2 \right\rangle \approx \left| e \right\rangle$ to $\left| spb2 \right\rangle \approx \left| b \right\rangle$ at the time $t_2 = \pi/(E_{2+} - E_{2-})$.

	We denote the probability of successful evolution of this stage as $P_2$ and choose $\gamma_{2} = 2/M + \epsilon_{2}$. When $M=100$, $1000$, and $10000$, to ensure $P_2 \geqslant 50\%$, the ranges of $\epsilon_{2}$ are $[-1.5 \times 10^{-3}, 1.5 \times 10^{-3}]$, $[-5.5 \times 10^{-3}, 5.5 \times 10^{-5}]$ and $[-1.9 \times 10^{-6}, 1.9 \times 10^{-6}]$, respectively. The detailed results are shown in Table \ref{gamma2Table}.
	
	\begin{table}
		\centering
		\caption{On the second-order lattice, when the target state is $\left| a \right\rangle$, the values of $\gamma_{2}$ are required in order to ensure $P_2 \geqslant 50\%$.}\label{gamma2Table}
		\begin{tabular}{cc p{15em}<{\centering} p{10em}<{\centering}}
			\hline
			$M$ & $\gamma_{c2}=2/M$ & $\gamma_{2}=\gamma_{c2} + \epsilon_2$ & $\frac{|\epsilon_2|}{\gamma_{c2}} \times 100\%$ \\
			\midrule[1pt]
			$100$ & $0.02$ & $1.85 \times 10^{-2} \sim 2.15 \times 10^{-2}$ & $7.5\%$ \\
			$1000$ & $0.002$ & $1.945 \times 10^{-3} \sim 2.055 \times 10^{-3}$ & $2.75\%$ \\
			$10000$ & $0.0002$ & $1.981 \times 10^{-4} \sim 2.019 \times 10^{-4}$ & $0.95\%$ \\
			\hline
		\end{tabular}
	\end{table}
	
	In the third stage of the algorithm, the initial state is $\left| b \right\rangle$, and the target state is $\left| a \right\rangle$. AS presented in Fig. \ref{H0_3}, in the subspace spanned by the basis states $|a\rangle$ and $|b\rangle$, the leading-order term $H_{3}^{(0)}=-\left(|a\rangle\langle a| +\gamma M |b\rangle\langle b|\right)$. We obtained the critical jumping rate $\gamma_{c3} = 1/M$ by setting the two eigenvalues to be equal.
	By introducing the perturbation $H_{3}^{(1)}=-\left(\gamma \sqrt{M}|a\rangle\langle b| +\gamma \sqrt{M} |b\rangle\langle a|\right)$, we obtain the eigenstates and eigenvalues of the whole Hamiltonian $H_3=H_3^{(0)} + H_3^{(1)}$

	\begin{align}
	|\phi_{3-}\rangle=\frac{1}{\sqrt{2}} (\left| a \right\rangle + \left| b \right\rangle) & , E_{3-}=-1-\sqrt{\frac{1}{M}},  \\
	|\phi_{3+}\rangle=\frac{1}{\sqrt{2}} (\left| a \right\rangle - \left| b \right\rangle) & , E_{3+}=-1+\sqrt{\frac{1}{M}}.
	\end{align}
	The system evolves into $\left| a \right\rangle$ when $t_3 = \pi / (E_{3+} - E_{3-})=\frac{\pi \sqrt{M}}{2}$.
	
	Assuming the success probability of this stage as $P_3$, we consider $\gamma_{3}=1/M+\epsilon_3$. To ensure $P_3 \geqslant 50\%$ when $M=100$, $1000$, and $10000$, $\epsilon_3$ must fall in the intervals of $[-2.5 \times 10^{-3}, 2.5 \times 10^{-3}]$, $[-5.5 \times 10^{-5}, 5.5 \times 10^{-5}]$, and $[-1.9 \times 10^{-6}, 1.9 \times 10^{-6}]$, respectively.
	
	\begin{table}
		\centering
		\caption{
		On the second-order lattice, when the target state is $\left| a \right\rangle$, the values of $\gamma_{3}$ are required in order to ensure $P_3 \geqslant 50\%$.}\label{gamma3Table}
		\begin{tabular}{cc p{15em}<{\centering} p{10em}<{\centering}}
			\hline
			$M$ & $\gamma_{c3}=1/M$ & $\gamma_{c}=\gamma_{c3}+\epsilon_3$ & $\frac{|\epsilon_3|}{\gamma_{c3}} \times 100\%$ \\
			\midrule[1pt]
			$100$ & $0.01$ & $0.75 \times 10^{-2} \sim 1.25 \times 10^{-2}$ & $25\%$ \\
			$1000$ & $0.001$ & $0.945 \times 10^{-3} \sim 1.055 \times 10^{-3}$ & $5.5\%$ \\
			$10000$ & $0.0001$ & $0.981 \times 10^{-4} \sim 1.019 \times 10^{-4}$ & $1.9\%$ \\
			\hline
		\end{tabular}
	\end{table}
	
\begin{figure}
		\centering
		\includegraphics[scale=0.5]{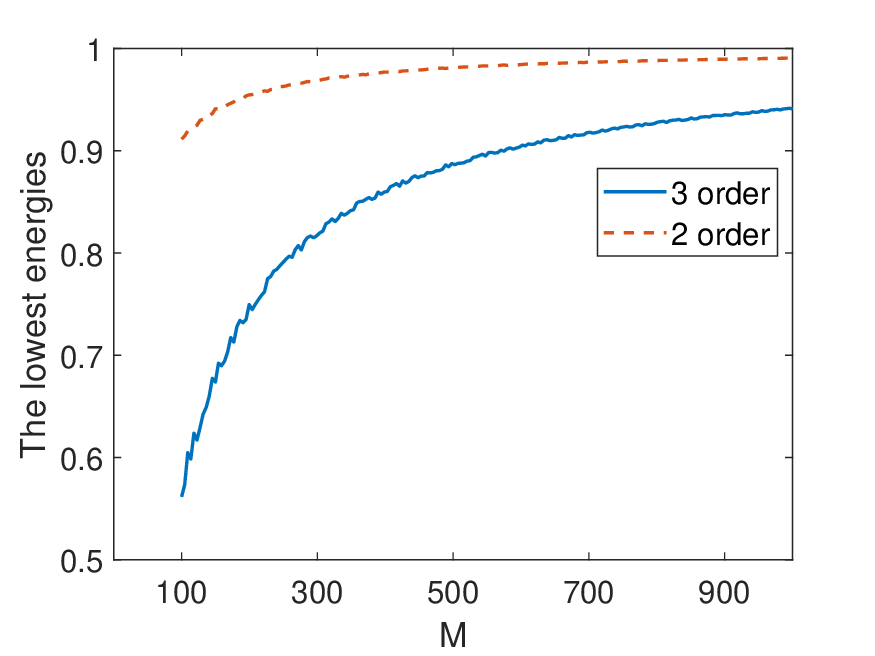}
		\caption{Success probabilities of the quantum search on the second- and third-order truncated M-simplex lattices.}\label{twoordersprobability}
	\end{figure}

	For the three-stage quantum search on the second-order lattice, we have obtained the three critical jumping rates $\gamma_{c1} = 3/M$, $\gamma_{c2} = 2/M$ and $\gamma_{c3} = 1/M$. The probability of a successful search on the second-order lattice is depicted in Fig. \ref{twoordersprobability}. As $M$ increases, the success probability is close to $100\%$.
	
	\begin{figure}[htbp]
		\centering
		\begin{subfigure}{0.9\linewidth}
			\centering
			\includegraphics[width=0.9\linewidth]{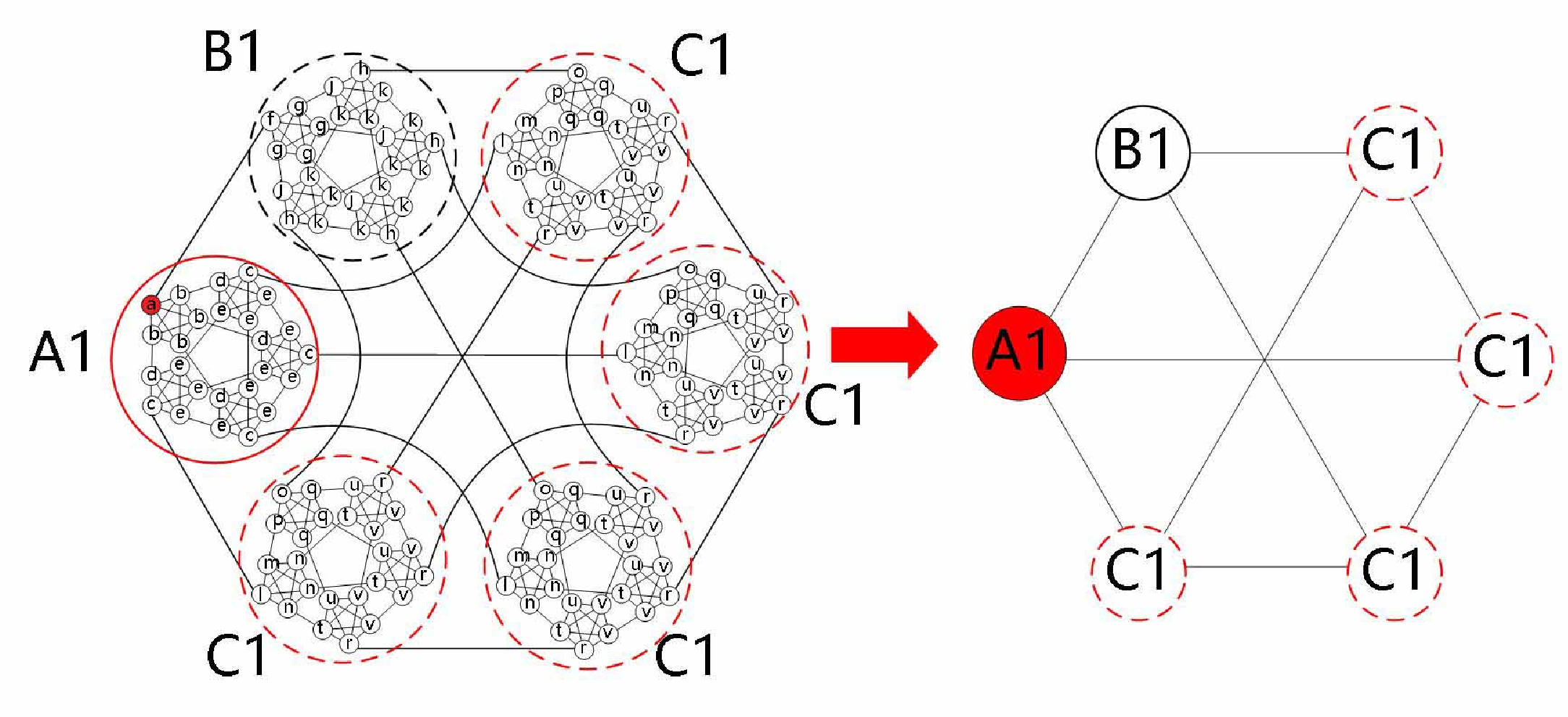}
			\caption{}
			\label{secondordermind1}
		\end{subfigure}
		\centering
		\begin{subfigure}{0.9\linewidth}
			\centering
			\includegraphics[width=0.9\linewidth]{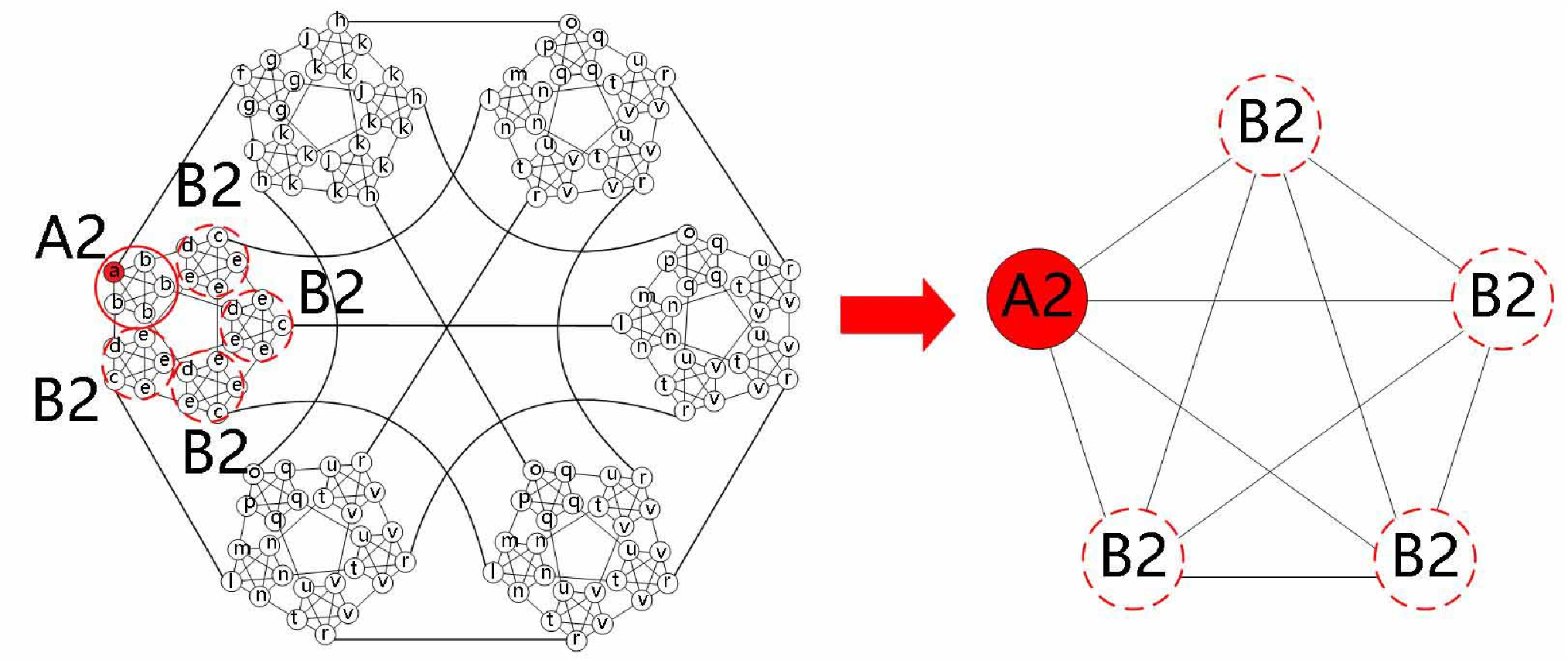}
			\caption{}
			\label{secondordermind2}
		\end{subfigure}
		\centering
		\begin{subfigure}{0.9\linewidth}
			\centering
			\includegraphics[width=0.9\linewidth]{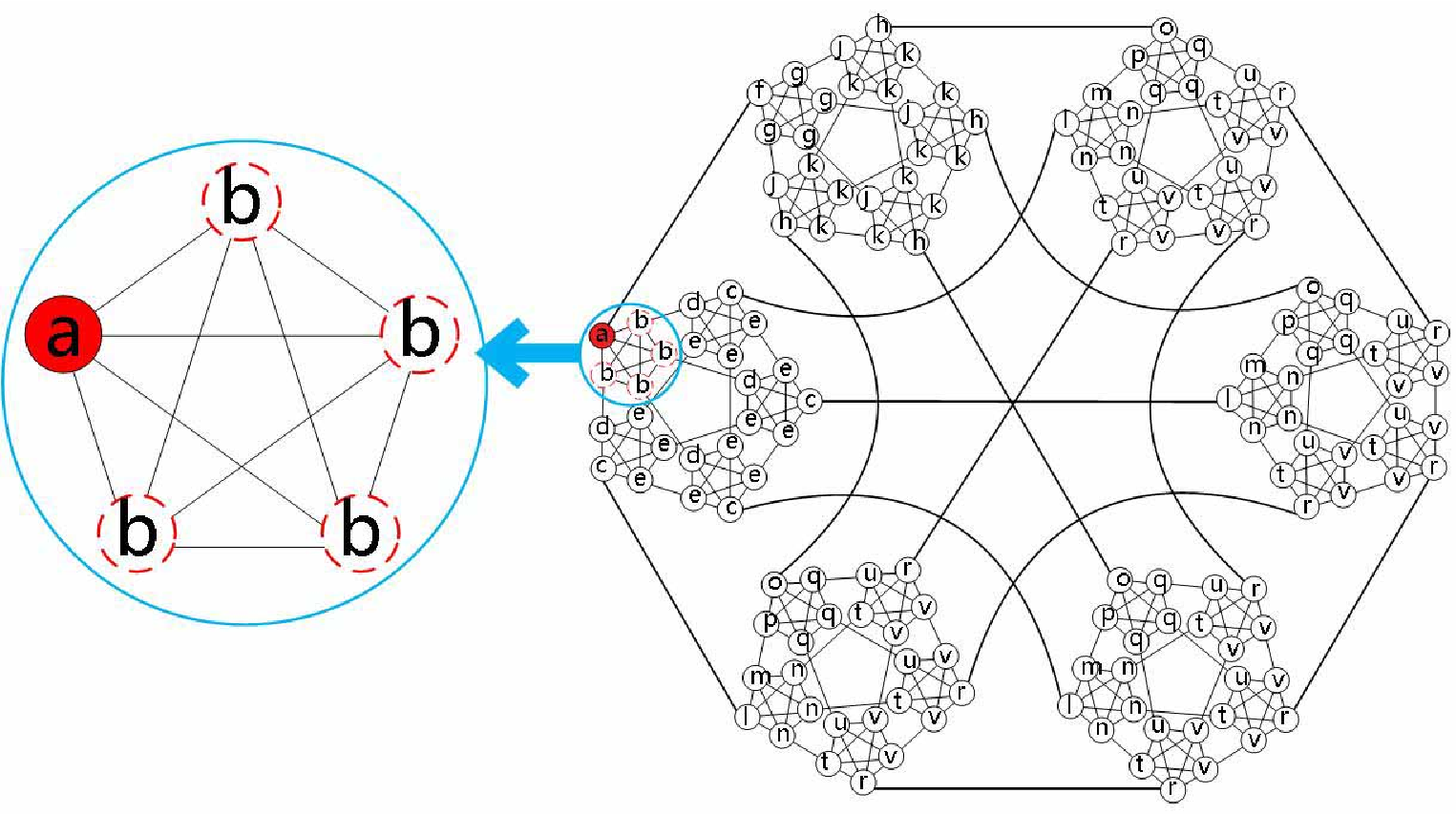}
			\caption{}
			\label{secondordermind3}
		\end{subfigure}
		\caption{Degenerate perturbation theory on second-order lattices.}
		\label{secondordermind}
	\end{figure}

\section{Rules for using degenerate perturbation theory}
	\label{sec:rules}
	In the second-order lattice, the structure formed by the set $\{\left| a \right\rangle, \left| b \right\rangle, \left| c \right\rangle, \left| d \right\rangle, \left| e \right\rangle\}$ is referred to as a first-order complete subgraph, which is denoted as $A1$ in Fig. \ref{secondordermind1}. Similarly, the structure formed by the set $\{\left| l \right\rangle, \left| m \right\rangle, \left| n \right\rangle, \left| o \right\rangle, \left| p \right\rangle, \left| q \right\rangle, \left| r \right\rangle, \left| t \right\rangle, \left| u \right\rangle, \left| v \right\rangle\}$ is also a first-order complete subgraph denoted by $C1$. It can be observed that the second-order lattice is constructed by $M+1$ first-order complete subgraphs, as depicted in Fig. \ref{secondordermind1}. $A1$, $B1$, and $C1$ represent distinct first-order complete subgraphs.
	
	The first stage of the quantum search on the second-order lattice takes place between the first-order complete subgraphs, from $C1$ to $A1$.
	We can omit the vertices in the first-order complete subgraph $B1$, and this has no influence. Therefore, we determined the critical jumping rate $\gamma_{c1}=3/M$ by considering only the two first-order complete subgraphs $C1$ and $A1$ to account for degeneracy.
		
	In the first stage of the search on the second-order lattice, there are two differences from the approach proposed in zeroth- and first-order lattice search. First, because the two sets $\{\left| a \right\rangle, \left| b \right\rangle\}$ and $\{\left| r \right\rangle, \left| t \right\rangle, \left| u \right\rangle, \left| v \right\rangle\}$ are not directly connected, the critical jumping rate $\gamma_{c1}$ cannot be determined by considering only these two subsystems.
	Second, the elements with a value of $1$ in the leading-order Hamiltonian cannot be omitted, and the approximations $M-l \approx M$ and $\sqrt{M-l} \approx \sqrt{M}$ are unavailable.
	The reason is that these approximations result in a significant change in the graph structure represented by the resulting matrix, compared to the original structure. In zeroth- and first-order structures, these changes are minor. In second-order and even higher-order structures, these changes are so substantial that the critical jumping rate cannot be accurately determined with these approximations.
		
	In the first-order complete subgraph $A1$, the structures that formed by the vertices $\{\left| a \right\rangle, \left| b \right\rangle\}$ and $\{\left| c \right\rangle, \left| d \right\rangle, \left| e \right\rangle\}$ are referred to as zeroth-order complete subgraphs, as depicted in Fig. \ref{secondordermind2}. $A2$ and $B2$ represent two categories of zeroth-order complete subgraphs.
	
	In the second stage of the quantum search, the evolution from $\left| e \right\rangle$ to $\left| b \right\rangle$ can be approximated as the evolution from $B2$ to $A2$. This stage occurs between two zeroth-order complete subgraphs. The third search stage from $\left| b \right\rangle$ to $\left| a \right\rangle$ occurs between two vertices in the zeroth-order complete subgraph $A2$, as shown in Fig. \ref{secondordermind3}.
	The structures involved in these two stages are simpler. Regardless of whether we retain the weighted edge $1$ and make the approximations $M-l \approx M$ and $\sqrt{M-l} \approx \sqrt{M}$, we can obtain the exact value of $\gamma_c$.
		
\begin{figure}
		\centering
		\includegraphics[scale=0.45]{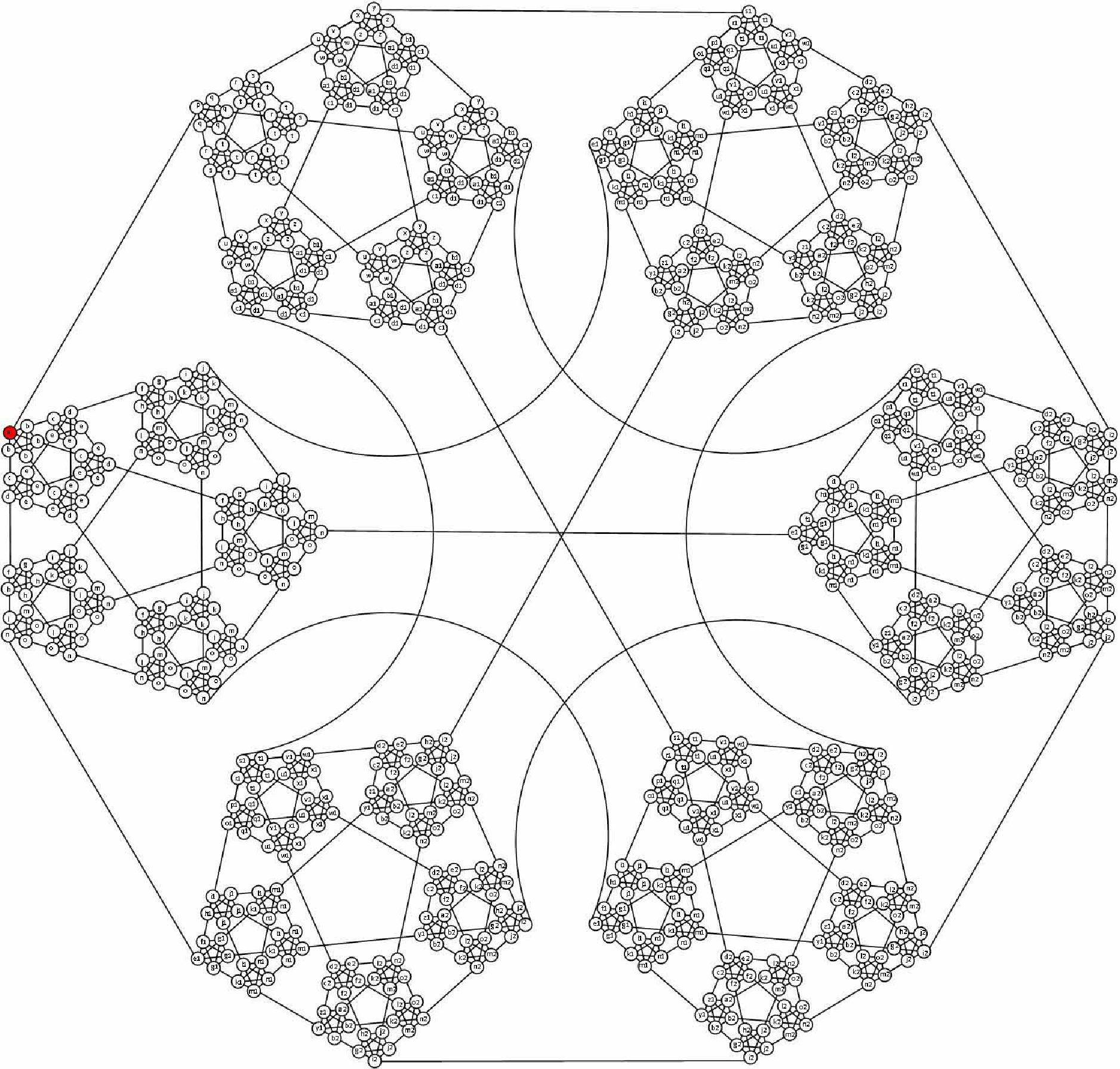}
		\caption{A third-order truncated five-dimensional simplex lattice. Vertices that evolving identically are labelled with the same letter. The red vertex is the marked vertex $a$.}\label{third-order}
	\end{figure}

	To prove the generality of the above method, we have extended the application of degenerate perturbation theory to the third-order lattice. 
	A third-order lattice is constructed by replacing each vertex of a second-order lattice with an $M$-dimensional complete graph, as depicted in Fig. \ref{third-order}. By symmetry, the system of the third-order lattice evolves in a $67$-dimensional invariant subspace. This subspace comprises the states $\left| a \right\rangle$, $\ldots$, $\left| z \right\rangle$, $\left| a1 \right\rangle$, $\ldots$, $\left| z1 \right\rangle$ and $\left| a2 \right\rangle$, $\ldots$, $\left| o2 \right\rangle$, as shown in Fig. \ref{third-order}.
	
\begin{figure}[htbp]
		\centering
		\begin{subfigure}{1\linewidth}
			\centering
			\includegraphics[width=0.55\linewidth]{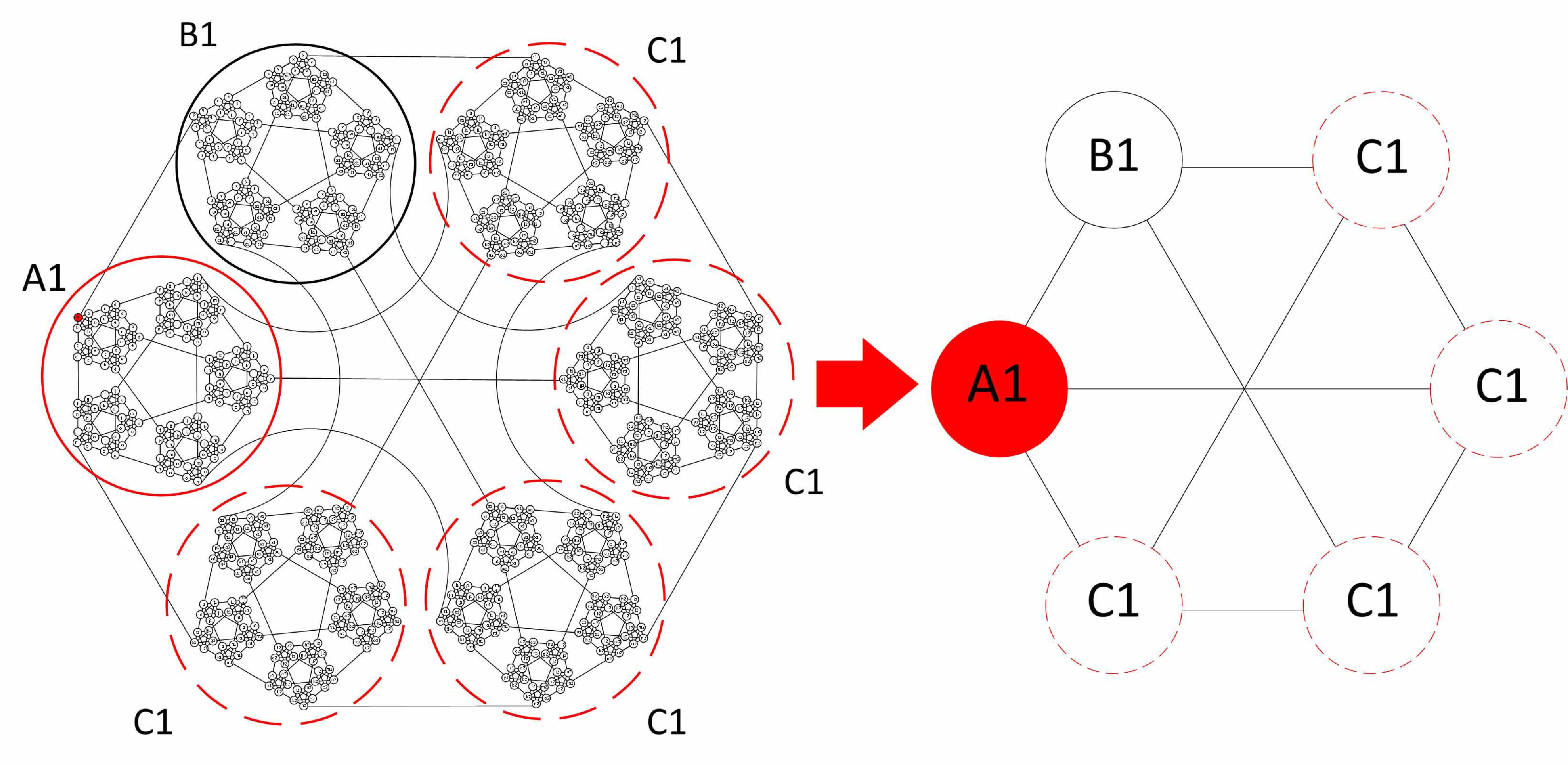}
			\caption{}
			\label{thirdordermind1}
		\end{subfigure}
		\centering
		\begin{subfigure}{1\linewidth}
			\centering
			\includegraphics[width=0.45\linewidth]{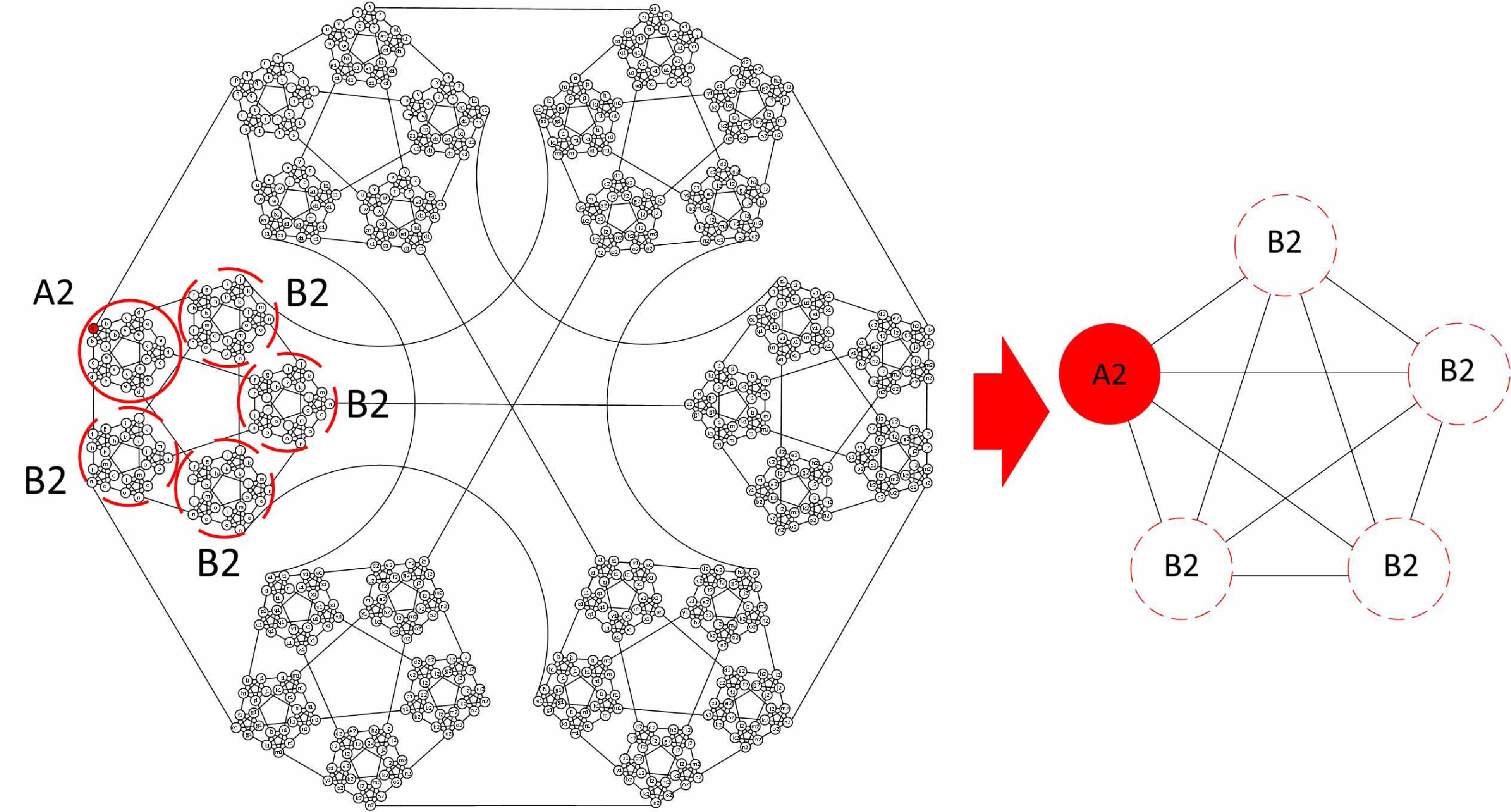}
			\caption{}
			\label{thirdordermind2}
		\end{subfigure}
		\centering
		\begin{subfigure}{1\linewidth}
			\centering
			\includegraphics[width=0.45\linewidth]{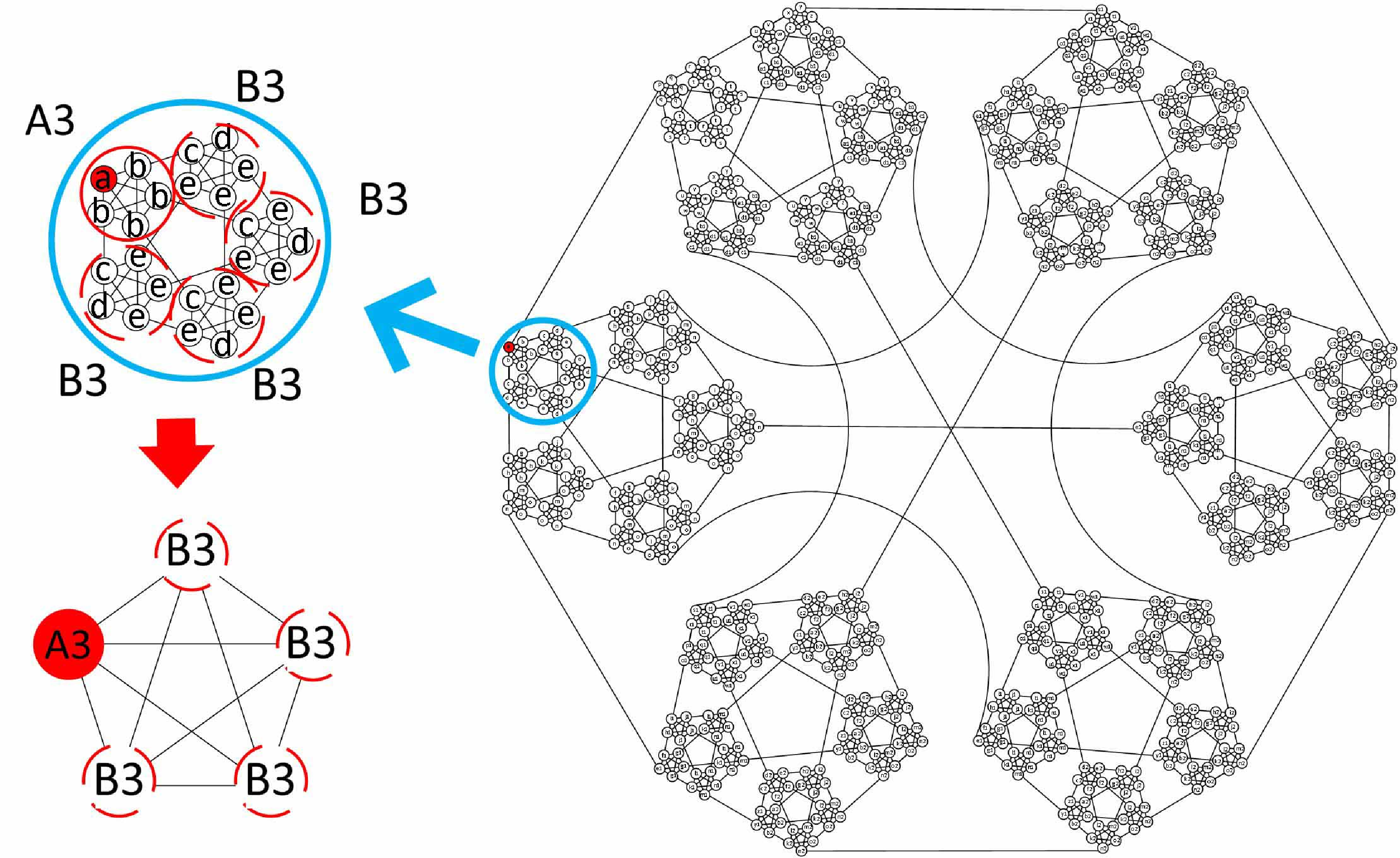}
			\caption{}
			\label{thirdordermind3}
		\end{subfigure}
		\centering
		\begin{subfigure}{1\linewidth}
			\centering
			\includegraphics[width=0.45\linewidth]{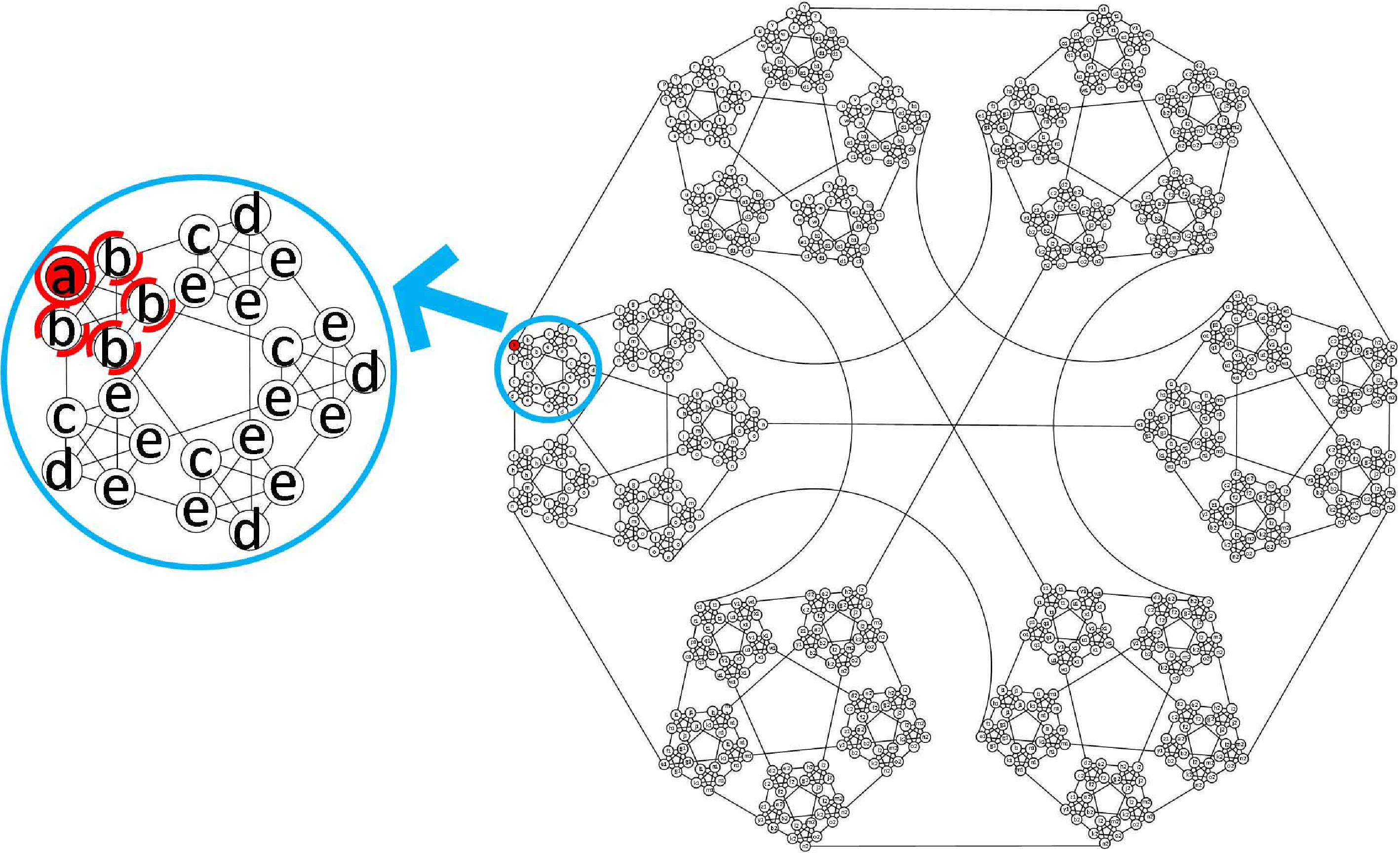}
			\caption{}
			\label{thirdordermind4}
		\end{subfigure}
		\caption{Degenerate perturbation theory on third-order lattice.}
		\label{thirdorderminds}
	\end{figure}

	The structures formed by the vertices corresponding to the sets $\{\left| a \right\rangle, \dots, \left| e \right\rangle\}$, $\{\left| p \right\rangle, \dots, \left| d1 \right\rangle\}$ and $\{\left| e1 \right\rangle, \dots, \left| o2 \right\rangle\}$ are referred to as second-order complete subgraphs.
	And the third-order lattice can be regarded as $M+1$ interconnected second-order complete subgraphs, as depicted in Fig. \ref{thirdordermind1}. $A1$, $B1$, and $C1$  represent three types of second-order complete subgraphs.

	The quantum search on the third-order lattice is a four-stage algorithm.
	When $M$ is large, the first stage of the quantum search on the third-order lattice can be seen as the evolution from the second-order complete subgraph $C1$ to $A1$. Hence, we consider $A1$ and $C1$ to account for degeneracy and determine the critical jumping rate $\gamma_{c1,3rd-order}$.
	
\begin{figure}[htbp]
		\centering
		\begin{subfigure}{0.4\linewidth}
			\centering
			\includegraphics[width=1\linewidth]{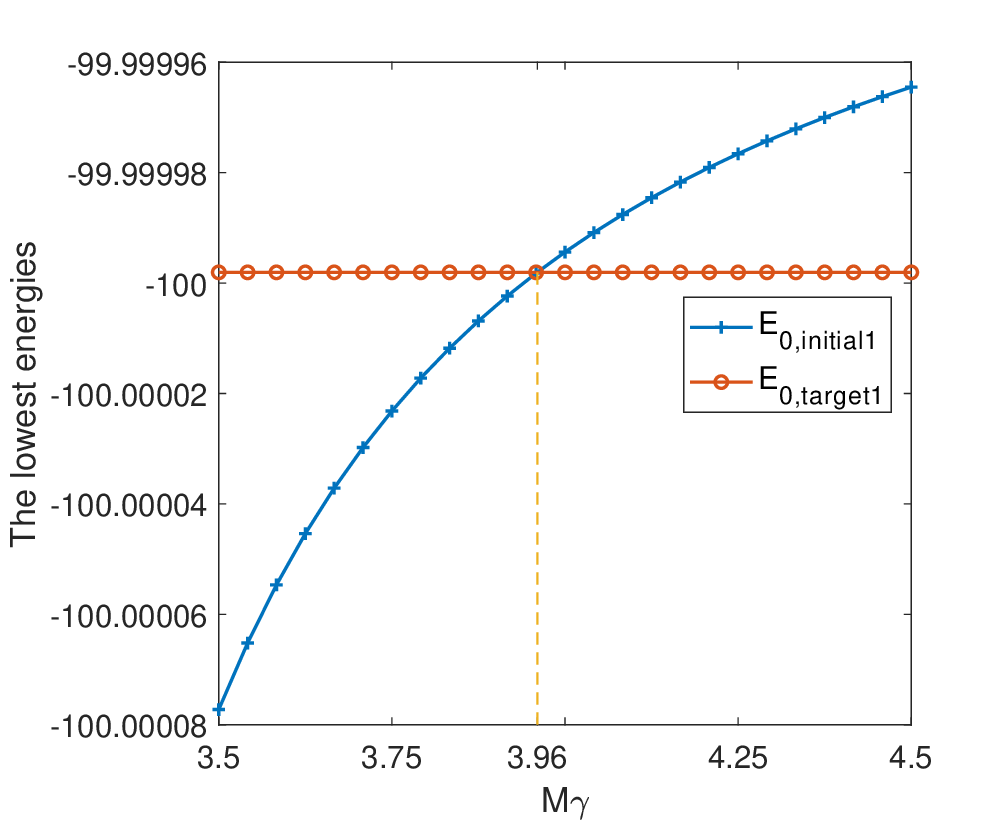}
			\caption{}
			\label{gamma1ofthirdorder}
		\end{subfigure}
		\centering
		\begin{subfigure}{0.4\linewidth}
			\centering
			\includegraphics[width=1\linewidth]{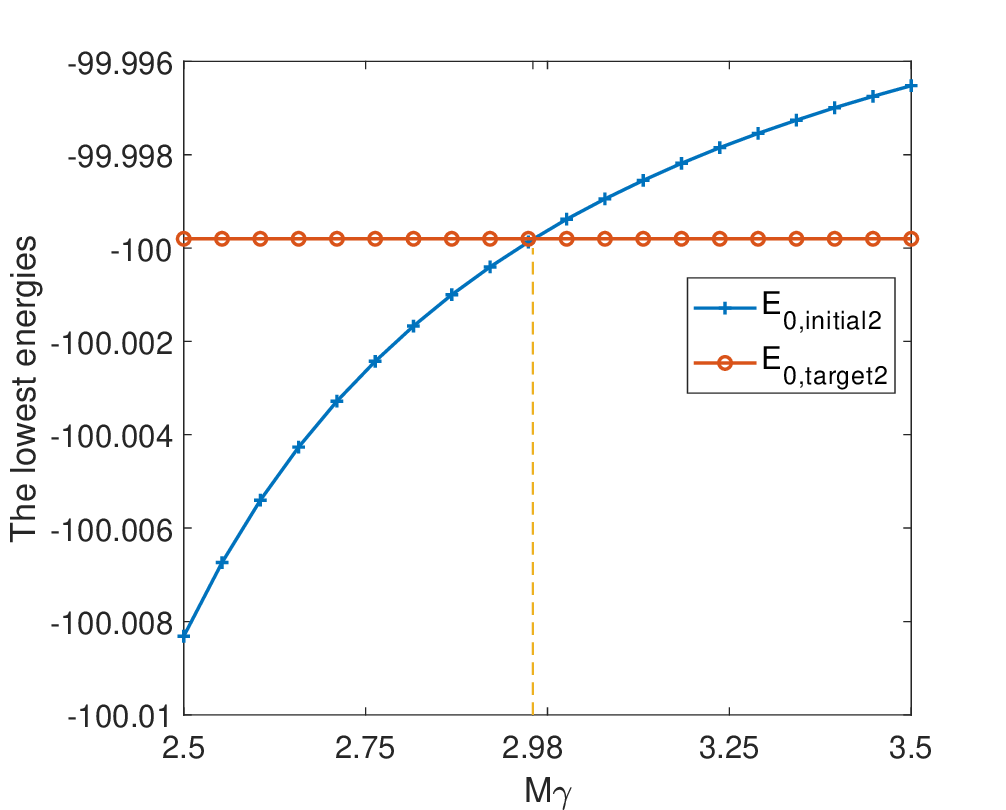}
			\caption{}
			\label{gamma2ofthirdorder}
		\end{subfigure}
		\begin{subfigure}{0.4\linewidth}
			\centering
			\includegraphics[width=1\linewidth]{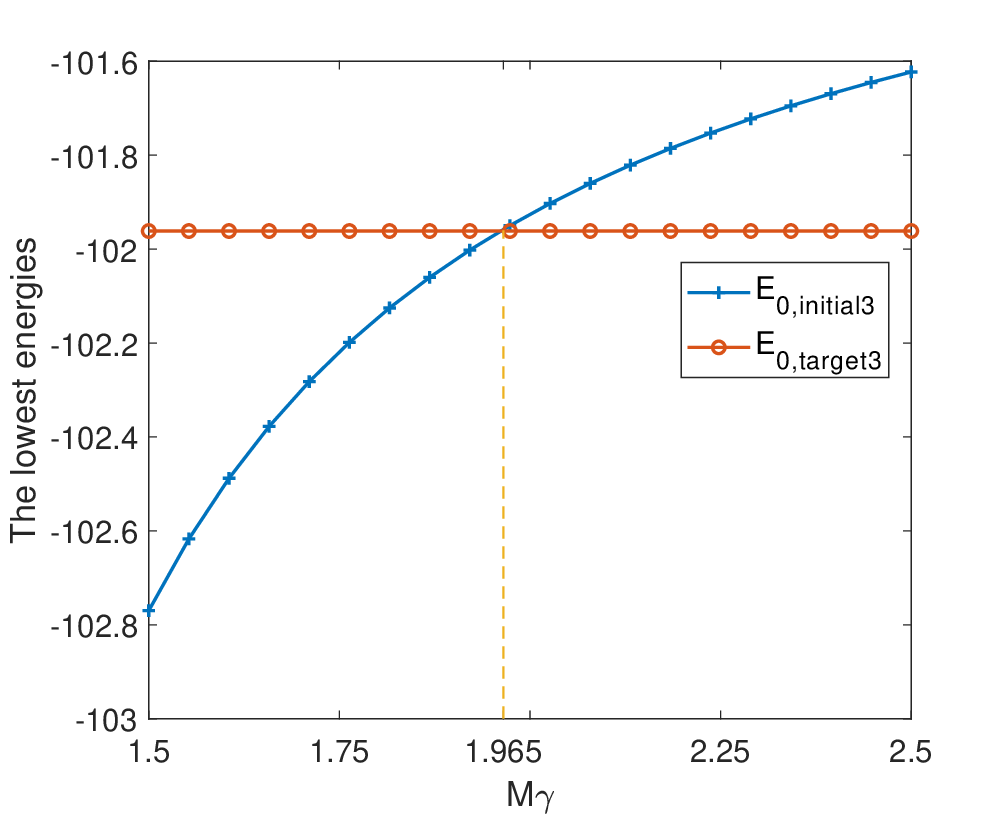}
			\caption{}
			\label{gamma3ofthirdorder}
		\end{subfigure}
		\caption{In quantum search on the third-order lattice, $\gamma_{c1,3rd-order} \approx 4/M$, $\gamma_{c2,3rd-order} \approx 3/M$, and $\gamma_{c3,3rd-order} \approx 2/M$ can be obtained through numerical calculations.}
		\label{gammasofthirdorder}
	\end{figure}

	Subsequently, the following stages of the quantum search follow a similar evolution as the one on the second-order lattice: at the second stage, the degeneracy of the first-order complete subgraphs represented by $A2$ and $B2$ in $A1$ are considered to determine $\gamma_{c2,3rd-order}$, as depicted in Fig. \ref{thirdordermind2}; at the third stage, we consider the degeneracy of the zeroth-order complete subgraphs represented by $A3$ and $B3$ in $A2$ in order to determine $\gamma_{c3,3rd-order}$, as shown in Fig. \ref{thirdordermind3}; at the fourth stage, we consider the degeneracy of the vertices $a$ and $b$ in $A3$ to determine $\gamma_{c4,3rd-order}$, as illustrated in Fig. \ref{thirdordermind4}.
	The critical jumping rates $\gamma_{c}$ of the three stages as depicted in Fig. \ref{gammasofthirdorder}: $\gamma_{c1,3rd-order} \approx 4/M$, $\gamma_{c2,3rd-order} \approx 3/M$, $\gamma_{c3,3rd-order} \approx 2/M$. For the fourth stage, $\gamma_{c4,3rd-order}=1/M$ can be directly computed as discussed in Section \ref{sec:secondorder}.
	
	It is important to note that at the first and second stages, the edges with a weight of $1$ in the second-order complete subgraphs and the first-order complete subgraphs can not be omitted. Additionally, the approximations $M-l \approx M$ and $\sqrt{M-l} \approx \sqrt{M}$ are not applicable.
	
\begin{figure}
		\centering
		\includegraphics[scale=0.1]{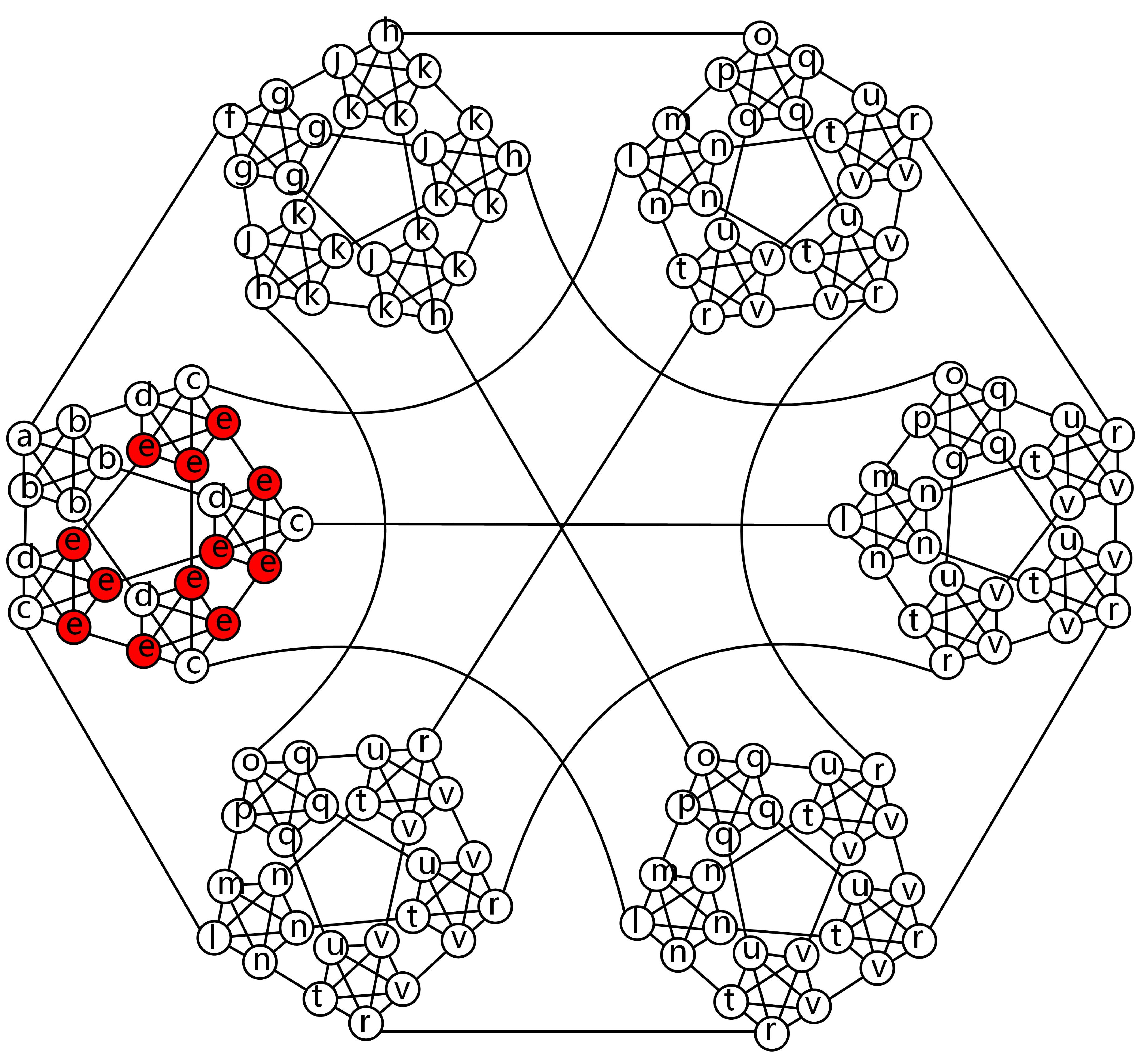}
		\caption{In the second-order lattice, we consider $\left| e \right\rangle$ as the marked state. The vertices that undergo the same evolution are represented by the same letter.}\label{second-ordermarkede}
	\end{figure}

	In the construction of leading-order terms based on different orders of complete subgraphs, we observed the correlation between the use of degenerate perturbation theory and the lattice structure.
	To demonstrate the dependency of degenerate perturbation theory on the structure, we have added two simulation experiments involving a set of marked vertices.
	For the second-order lattice, we select $\left| e \right\rangle$ as the marked state which contains $(M-1)(M-2)$ vertices, as depicted in Fig. \ref{second-ordermarkede}. The search Hamiltonian is
	\begin{equation}
		H=-\gamma A- \left| e \right\rangle\left\langle e \right|.
	\end{equation}
	
\begin{figure}[htbp]
		\begin{subfigure}{0.5\linewidth}
			\centering
			\includegraphics[width=0.9\linewidth]{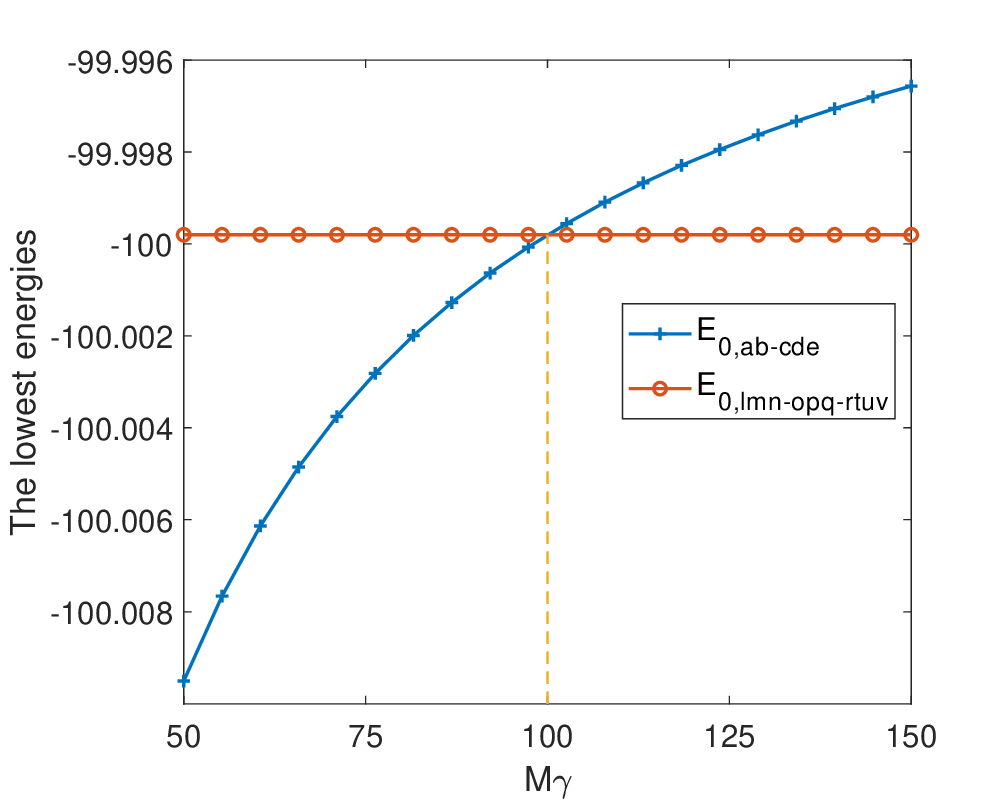}
			\caption{}
			\label{e_abcde}
		\end{subfigure}
		\begin{subfigure}{0.5\linewidth}
			\centering
			\includegraphics[width=0.9\linewidth]{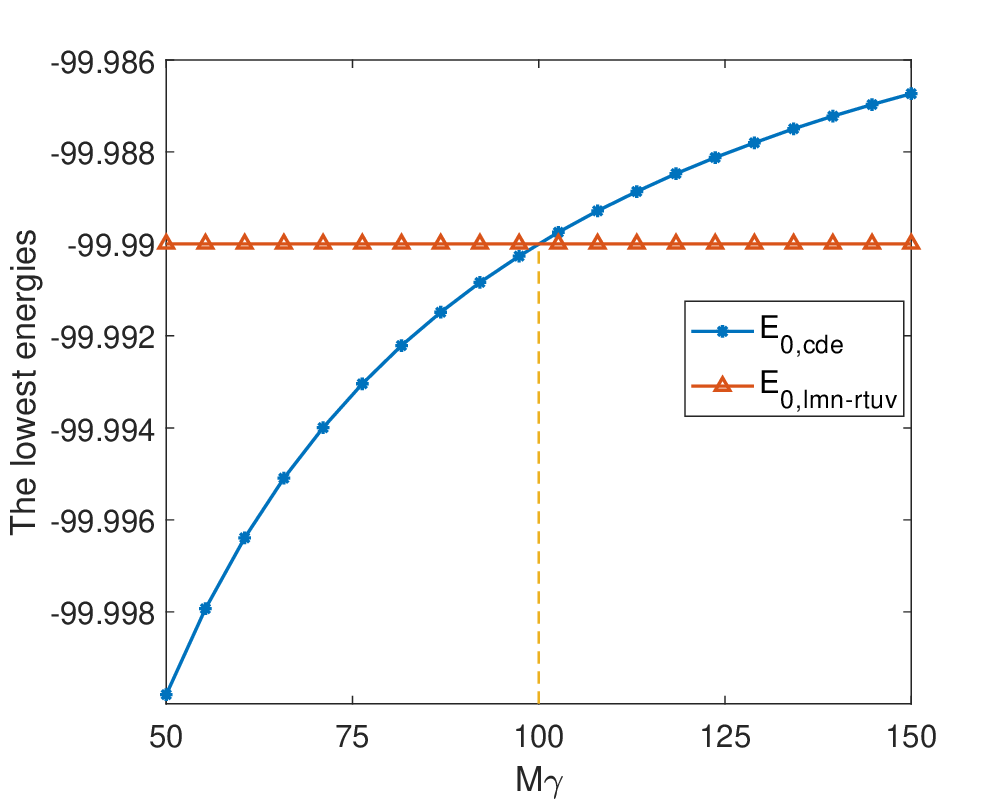}
			\caption{}
			\label{e_cde}
		\end{subfigure}
		\caption{In the second-order lattice, $M=100$, and $\left| e \right\rangle$ is the marked state. \textbf{a} We use   $\{\left| a \right\rangle,\left| b \right\rangle, \left| c \right\rangle,\left| d \right\rangle,\left| e \right\rangle\}$ and $\{\left| l \right\rangle,\left| m \right\rangle,\left| n \right\rangle, \left| o \right\rangle,\left| p \right\rangle,\left| q \right\rangle,\left| r \right\rangle,\left| t \right\rangle,\left| u \right\rangle,\left| v \right\rangle\}$ to construct the Hamiltonian.
			\textbf{b} We use $\{\left| c \right\rangle,\left| d \right\rangle,\left| e \right\rangle\}$ and $\{\left| l \right\rangle,\left| m \right\rangle,\left| n \right\rangle,\left| r \right\rangle,\left| t \right\rangle,\left| u \right\rangle,\left| v \right\rangle\}$ to construct the Hamiltonian.}
		\label{e}
	\end{figure}

	This quantum search is a one stage algorithm. Here, we use two schemes to determine the critical jumping rate.
	The first scheme closely resembles the one discussed in Section \ref{sec:secondorder}. The leading-order term of the Hamiltonian aligns with the one given by Eq. \ref{H0ofstage1}, differing only in
	\begin{equation}
		H_{ab-cde}^{(0)} = -\gamma \left(
		\begin{array}{ccccc}
			0 & \sqrt{M_1} & 0 & 0 & 0 \\
			\sqrt{M_1} & M_2 & 0 & 1 & 0 \\
			0 & 0 & 0 & 1 & \sqrt{M_2} \\
			0 & 1 & 1 & 0 & \sqrt{M_2} \\
			0 & 0 & \sqrt{M_2} & \sqrt{M_2} & \frac{1}{\gamma}+M_2\\
		\end{array} \right).
	\end{equation}
	The critical jumping rate $\gamma_{c,marked(e)}$ is determined by the two lowest energies $E_{0,ab-cde}$ and $E_{0,lmn-opq-rtuv}$ of the subsystems $\{\left| a \right\rangle,\left| b \right\rangle,\left| c \right\rangle,\left| d \right\rangle,\left| e \right\rangle\}$ and $\{\left| l \right\rangle,\left| m \right\rangle,\left| n \right\rangle,\left| o \right\rangle, \left| p \right\rangle, \left| q \right\rangle, \left| r \right\rangle, \left| t \right\rangle, \left| u \right\rangle, \left| v \right\rangle\}$, respectively. And $\gamma_{c,marked(e)}=M$ as illustrated in Fig. \ref{e_abcde}.
	

\begin{figure}
		\centering
		\includegraphics[scale=0.09]{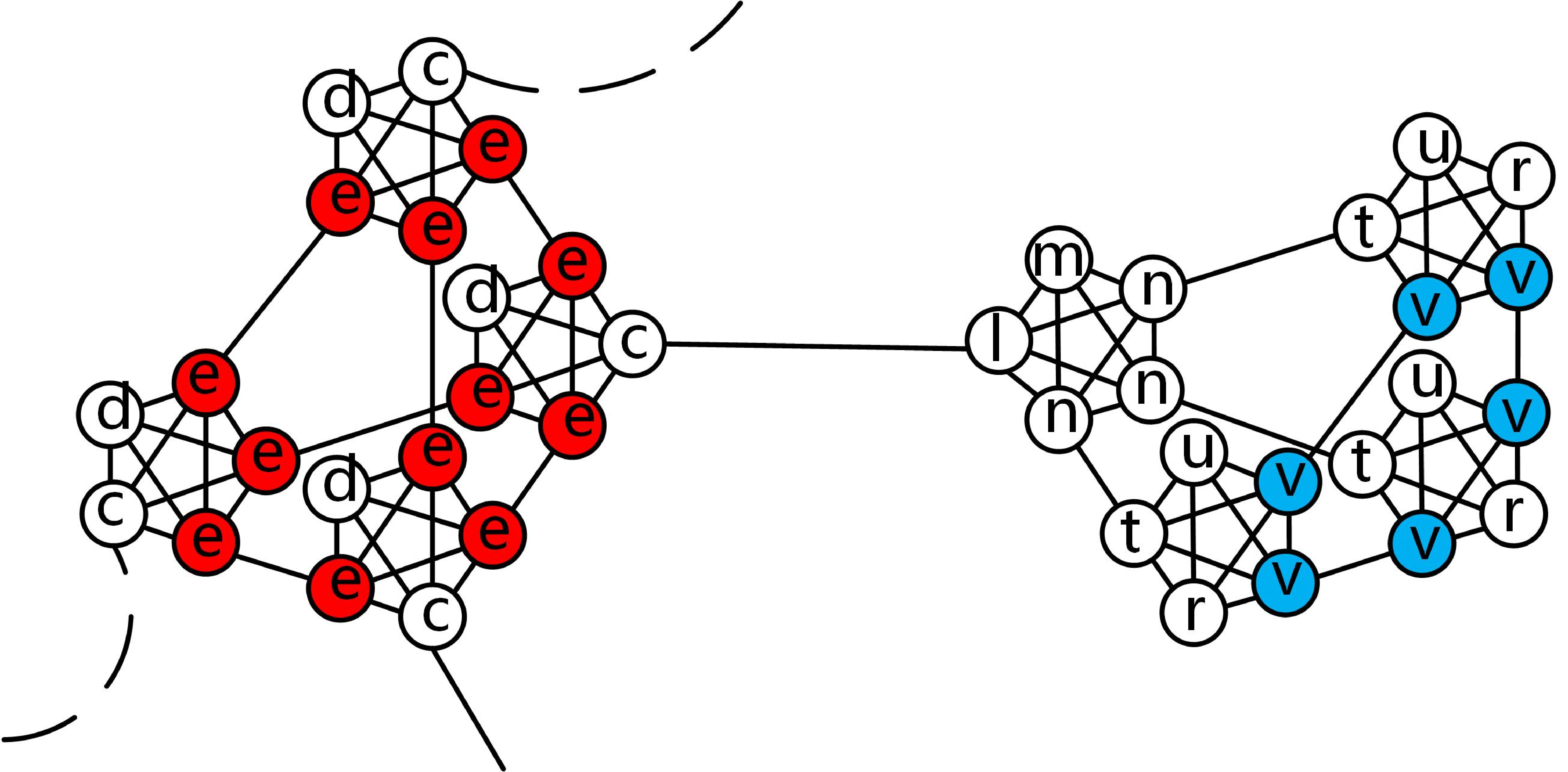}
		\caption{When considering $\left| e \right\rangle$ as the marked state, the leading-order term that only involves $\{\left| c \right\rangle,\left| d \right\rangle,\left| e \right\rangle\}$ and $\{\left| l \right\rangle,\left| m \right\rangle,\left| n \right\rangle,\left| r \right\rangle,\left| t \right\rangle,\left| u \right\rangle,\left| v \right\rangle\}$ is shown. The dashed lines from the vertices $c$ connect to the other structures formed by the set $\{\left| l \right\rangle,\left| m \right\rangle,\left| n \right\rangle,\left| r \right\rangle,\left| t \right\rangle,\left| u \right\rangle,\left| v \right\rangle\}$.}\label{e_structure}
	\end{figure}

	Based on the lattice structure shown in Fig. \ref{second-ordermarkede}, the set $\{\left| r \right\rangle,\left| t \right\rangle,\left| u \right\rangle,\left| v \right\rangle\}$ is connected to the set $\{\left| c \right\rangle,\left| d \right\rangle,\left| e \right\rangle\}$ through the set $\{\left| l \right\rangle,\left| m \right\rangle,\left| n \right\rangle\}$. The two sets $\{\left| a \right\rangle,\left| b \right\rangle\}$ and $\{\left| o \right\rangle,\left| p \right\rangle,\left| q \right\rangle\}$ are unrelated to the evolution.
	Therefore, we only retain the parts of $\{\left| c \right\rangle,\left| d \right\rangle,\left| e \right\rangle\}$ and $\{\left| l \right\rangle,\left| m \right\rangle\, \left| n \right\rangle,\left| r \right\rangle\, \left| t \right\rangle,\left| u \right\rangle,\left| v \right\rangle\}$ in the leading-order term, as shown in Fig. \ref{e_structure}. We consider this method as the second scheme.
	By setting the lowest energy of $\{\left| c \right\rangle,\left| d \right\rangle,\left| e \right\rangle\}$ equal to the lowest energy of $\{\left| l \right\rangle,\left| m \right\rangle\, \left| n \right\rangle,\left| r \right\rangle\, \left| t \right\rangle,\left| u \right\rangle,\left| v \right\rangle\}$,  the critical jumping rate $\gamma'_{c,marked(e)} = M$ can be determined, as illustrated in Fig. \ref{e_cde}. This result is consistent with the first scheme. Further calculations confirm that the system indeed has a high probability of evolving to $\left| e \right\rangle$ when $\gamma_{marked(e)}=M$.
	
	It is found that the success of the second scheme can be attributed to the fact that after omitting $\{\left| a \right\rangle,\left| b \right\rangle\}$ and $\{\left| o \right\rangle,\left| p \right\rangle,\left| q \right\rangle\}$, the structures corresponding to the sets $\{\left| c \right\rangle,\left| d \right\rangle,\left| e \right\rangle\}$ and $\{\left| l \right\rangle,\left| m \right\rangle,\left| n \right\rangle,\left| r \right\rangle,\left| t \right\rangle,\left| u \right\rangle,\left| v \right\rangle\}$ share the same overall shape, despite containing different types of vertices. The same overall shapes of the two subsystems are called structural consistency.
	
	Now, we summarize the guidelines of using degenerate perturbation theory for the quantum search on an $r$th-order lattice.
	Here, we provide several definitions:
	(1) The $(r-1)$th-order complete subgraph is defined as the secondary structure of an $r$th-order lattice, and similarly, a $(r-2)$th-order complete subgraph is considered  as the secondary structure of the $(r-1)$th-order complete subgraph.
	(2) We designate the secondary structure associated with the initial state as the "initial secondary structure" and the secondary structure associated with the target state as the "target secondary structure".
	(3) When some vertices within the $(r-1)$th-order complete subgraph are excluded, we still refer to the resulting structure as a secondary structure of the $r$th-order lattice.
	(4) The set of basis states associated with a complete graph is referred to as a basis group, like $\{\left| a \right\rangle,\left| b \right\rangle\}$ and $\{\left| o \right\rangle,\left| p \right\rangle,\left| q \right\rangle\}$.

	The first step of the quantum search is to identify the initial and target secondary structures. Afterward, we eliminate the basis groups unrelated to the search. Finally, we establish degeneracy between the remaining components of the two secondary structures to determine the critical jumping rate.
	A basis group can be considered as a set unrelated to the search and can be omitted only when it satisfies the following two constraints:
	(1) The basis group is not part of the shortest path between the initial and target states.
	(2)	After the omission of the basis groups, structural consistency can still be retained in the initial and target secondary structures.
	Following these two constraints, we can effectively reduce the dimension of the Hamiltonian through this omission, resulting in simplified computations.
	
\begin{figure}
		\centering
		\includegraphics[scale=0.4]{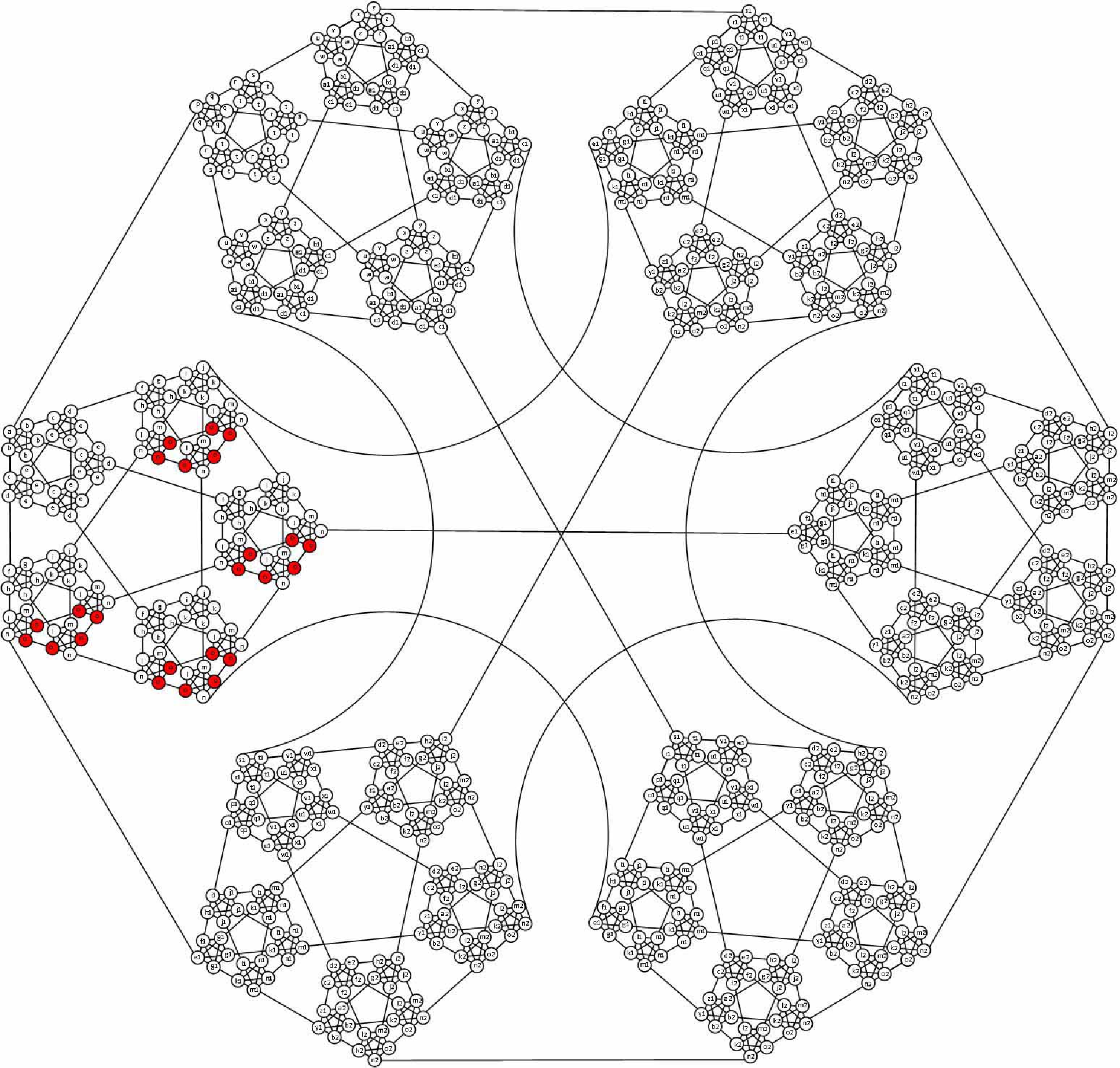}
		\caption{In the third-order truncated five-dimensional simplex lattice, $\left| o \right\rangle$ is the marked state.}\label{third-ordermarkedo} 
	\end{figure}

\begin{figure}[htbp]
		\centering
		\begin{subfigure}{0.4\linewidth}
			\centering
			\includegraphics[width=0.6\linewidth]{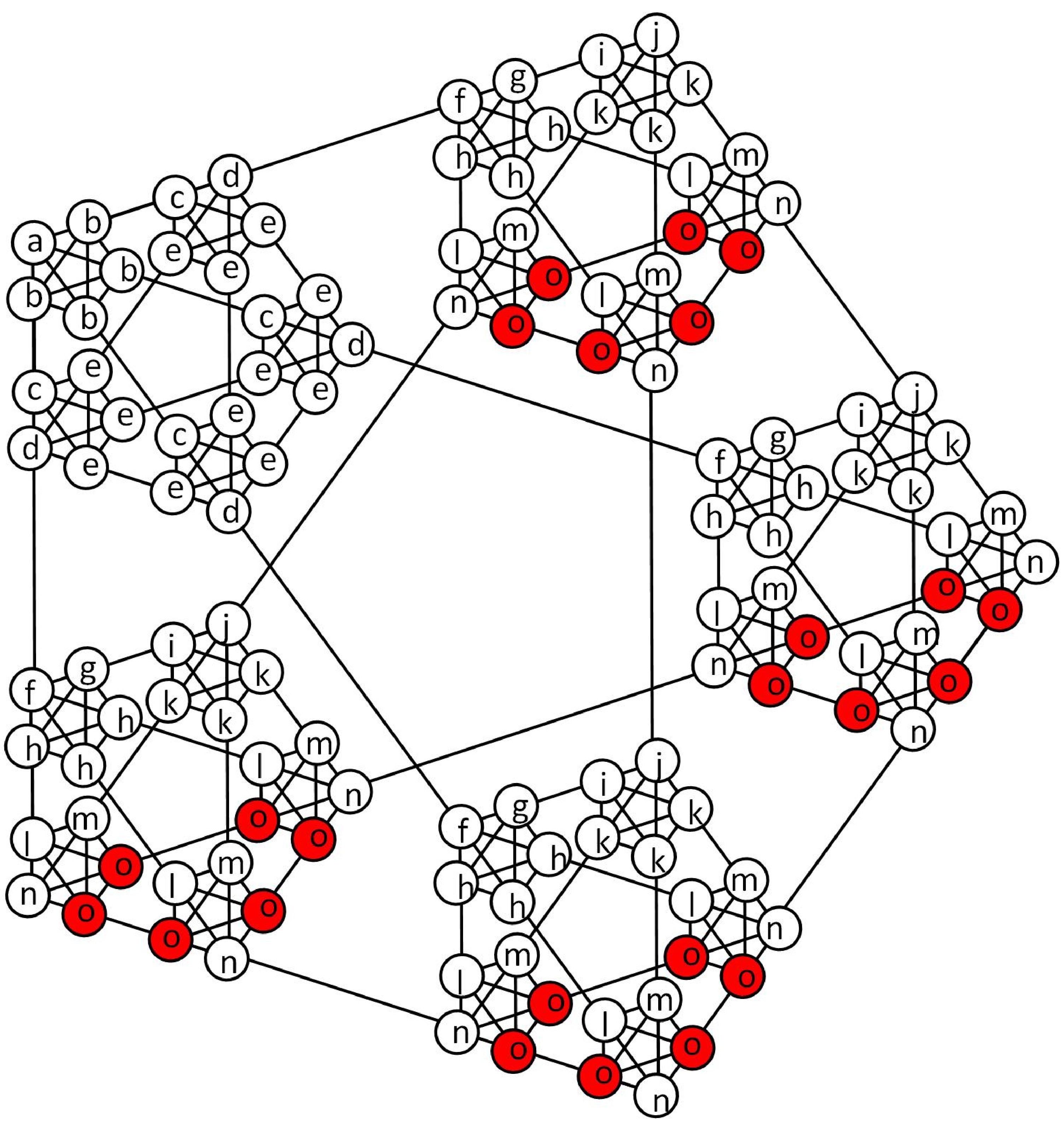}
			\caption{}
			\label{thirdorderofatoo}
		\end{subfigure}
		\centering
		\begin{subfigure}{0.4\linewidth}
			\centering
			\includegraphics[width=0.6\linewidth]{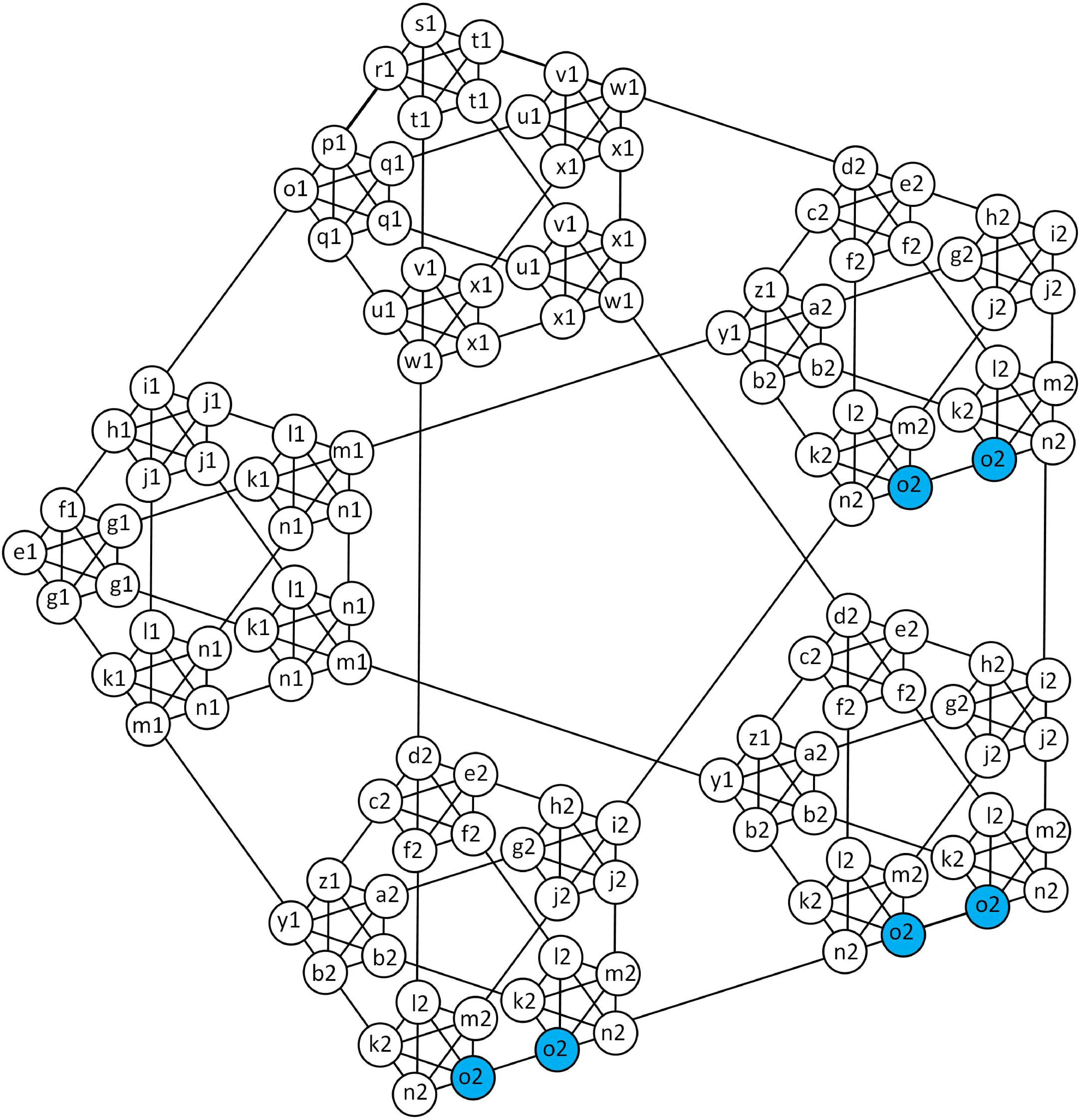}
			\caption{}
			\label{thirdorderofe1too2}
		\end{subfigure}
		\centering
		\begin{subfigure}{0.4\linewidth}
			\centering
			\includegraphics[width=0.6\linewidth]{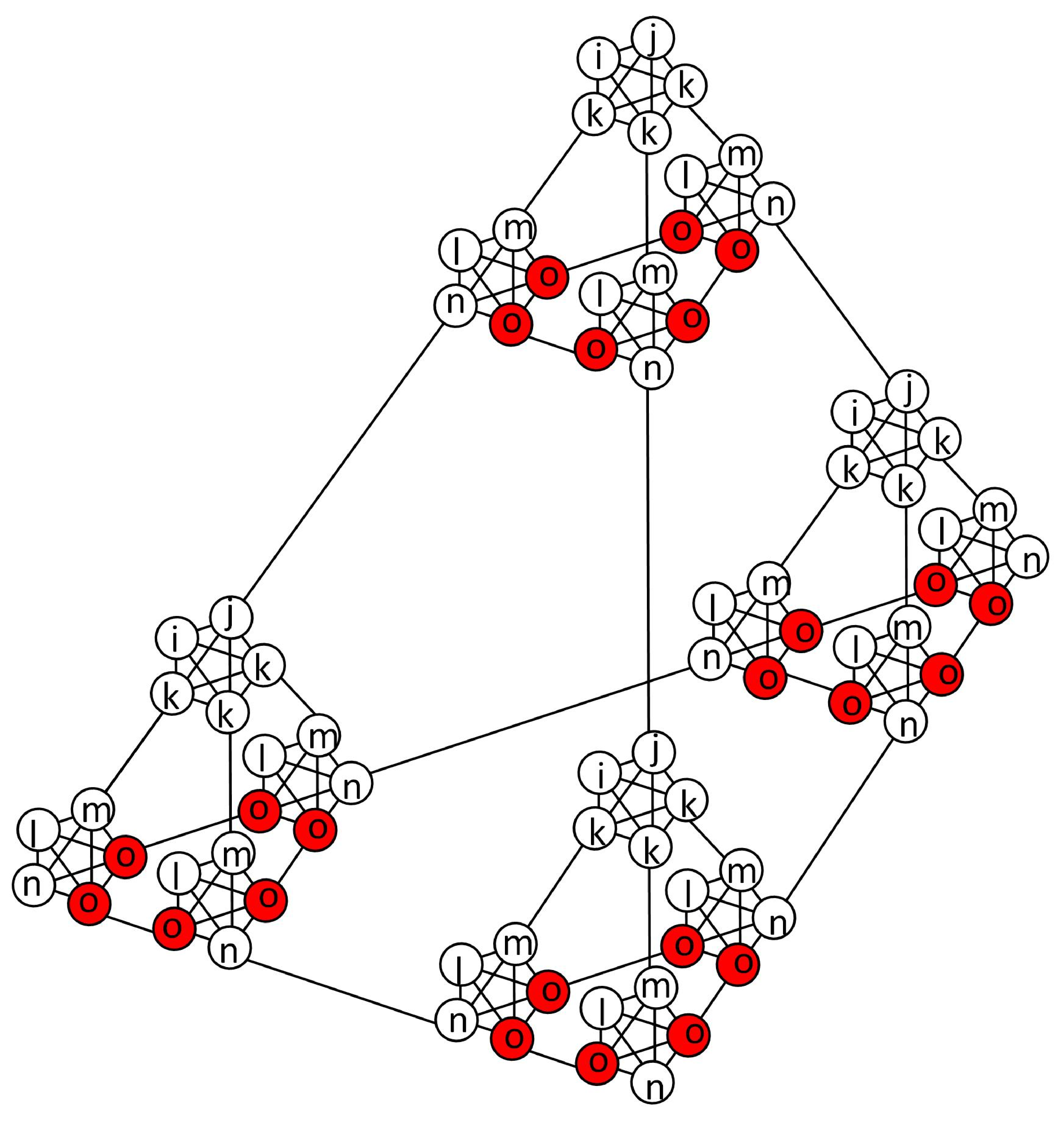}
			\caption{}
			\label{thirdorderofatoowithomition}
		\end{subfigure}
		\centering
		\begin{subfigure}{0.4\linewidth}
			\centering
			\includegraphics[width=0.6\linewidth]{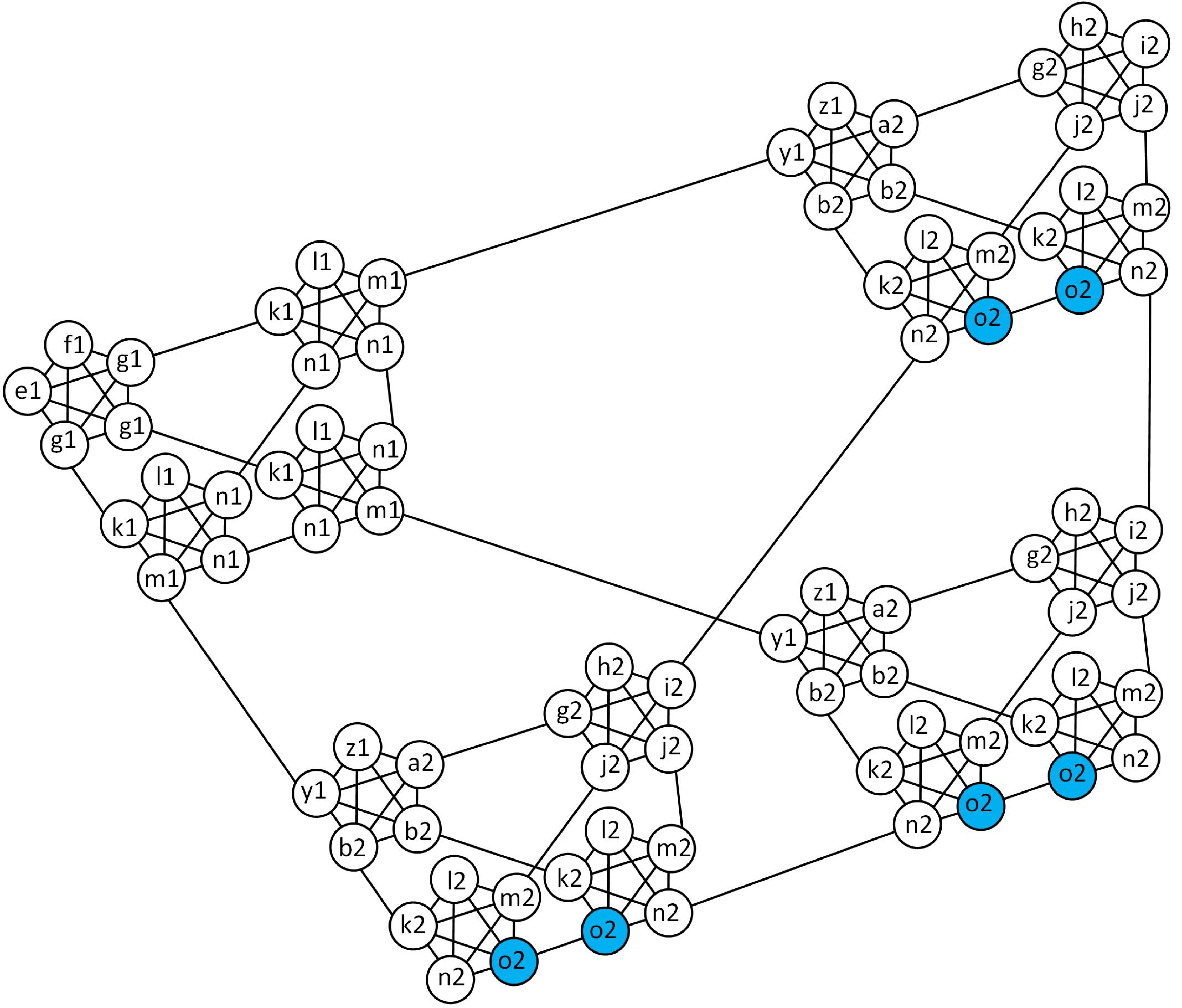}
			\caption{}
			\label{thirdorderofe1too2withomition}
		\end{subfigure}
		\caption{
			In the third-order lattice, when $\left| o \right\rangle$ is the marked state, we displayed the target and initial secondary structures, as well as the modified versions of these structures obtained after omitting specific basis groups that are unrelated to the search.
		}
		\label{thirdorderofdifferentmarkers}
	\end{figure}

	We consider the search on the third-order lattice with the marked state  $\left| o \right\rangle$ in Fig. \ref{third-ordermarkedo} as an example to demonstrate the above guidelines. The Hamiltonian of the system is
	\begin{equation}
		H=-\gamma A- \left| o \right\rangle\left\langle o \right|.
	\end{equation}
	We have determined the critical jumping rates by using the above two schemes, as illustrated in Fig. \ref{thirdorderofdifferentmarkers}. The two schemes all provide the correct critical jumping rate $M^2$.
	It has been demonstrated that when $\gamma_{marked(o)}=M^2$, the system can effectively evolve from the state $\left| o2 \right\rangle$ to the state $\left| o \right\rangle$ with a high probability.
	
\section{The impact of varying the positions of marked vertices}
	\label{sec:theinfluence}
	
	\begin{figure}
		\centering
		\includegraphics[scale=0.075]{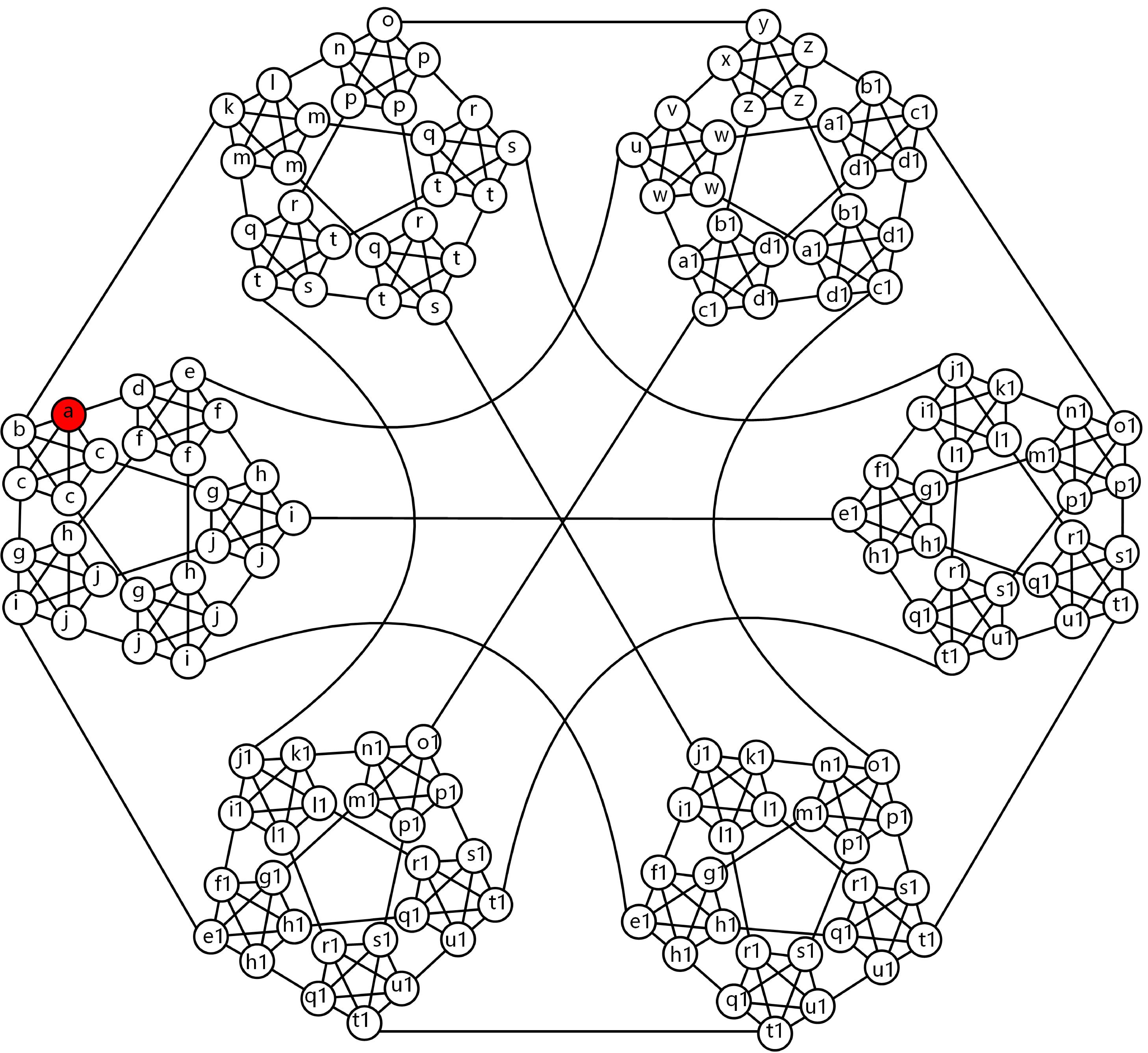}
		\caption{Another case where there's only one marked vertice on a second-order lattice.}\label{secondordercase2} 
	\end{figure}
	
	In the second-order lattice, the marked vertex labeled as $a$ can be positioned differently, as illustrated in Fig. \ref{secondordercase2}.
	This variation leads to an increase in the dimension of the invariant subspace from $20$ to $47$, which includes states $\{\left| a \right\rangle,\dots,\left| z \right\rangle\}$ and $\{\left| a1 \right\rangle,\dots,\left| u1 \right\rangle\}$.
	The initial state is the uniform superposition of all states, given by $\left| {{\psi}}(0) \right\rangle =\frac{1}{\sqrt{N}} \sum_{i=1}^{N} \left| i \right\rangle$. When $M$ is large, we have $\left| {{\psi}}(0) \right\rangle \approx \left| u1 \right\rangle$.
	
	\begin{figure}
		\centering
		\includegraphics[scale=0.4]{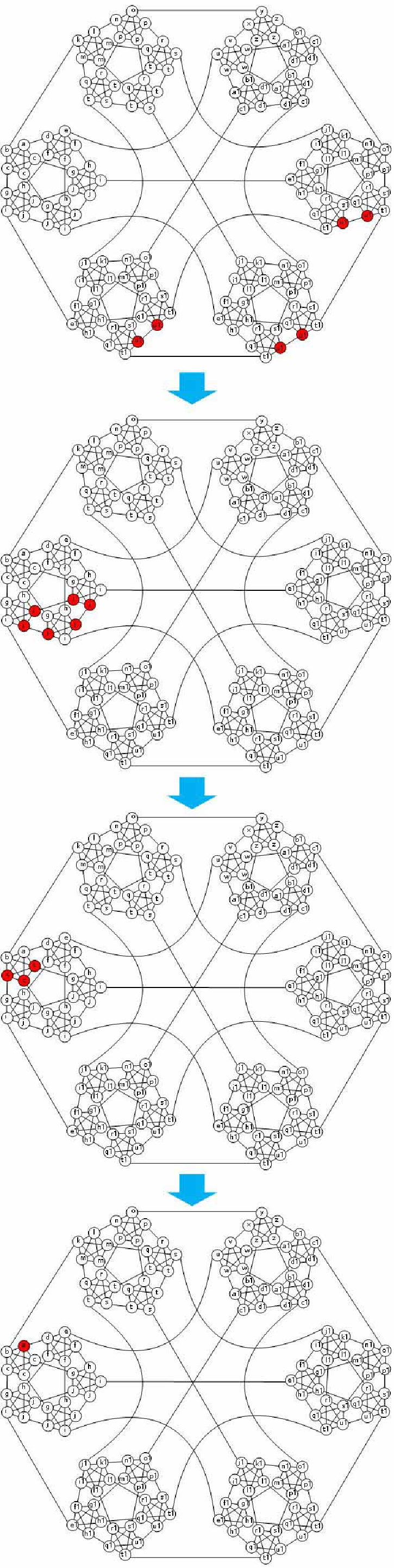}
		\caption{
			The evolution to the target state on the second lattice when the marked vertex is positioned differently.}\label{secondordercase2mind} 
	\end{figure}
	
	The evolution of the system still has three stages, from $\left| {{\psi}}(0) \right\rangle  \approx \left| u1 \right\rangle$ to $\left| j \right\rangle$ , then to $\left| c \right\rangle$, and finally to $\left| a \right\rangle$, as shown in Fig. \ref{secondordercase2mind}.
	The critical jumping rates of the three stages are $\gamma'_{c1}=3/M$, $\gamma'_{c2}=2/M$ and $\gamma'_{c3}=1/M$, respectively.
	This result is consistent with the findings in Section \ref{sec:secondorder}. It can be observed that in the second-order lattice, when there is one marked vertex, the different positions of the marked vertex does not affect the values of the critical jumping rates.
	
	For the two quantum searches on the second-order lattice where there is only one marked vertex, as shown in Figs. \ref{second-order} and \ref{secondordercase2}, further calculations reveal that the time required for each stage is approximately the same.
	The corresponding structures of the two different marked vertices for each stage are identical, which implies that the different positions of the marked vertex have no effect on the search.
	
\begin{figure}[htbp]
		\begin{subfigure}{0.5\linewidth}
			\centering
			\includegraphics[width=0.8\linewidth]{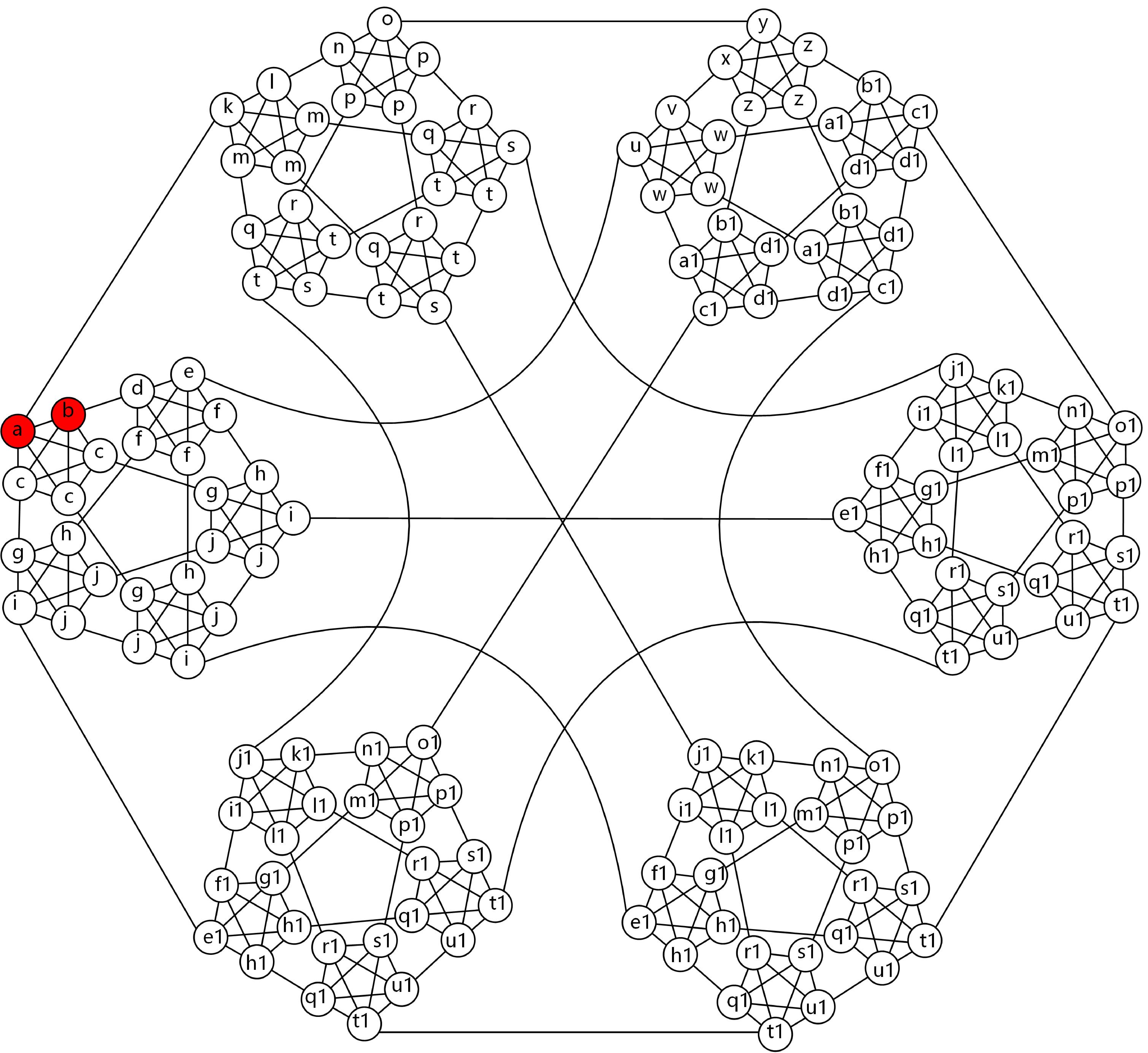}
			\caption{}
			\label{twomarkers1}
		\end{subfigure}
		\begin{subfigure}{0.5\linewidth}
			\centering
			\includegraphics[width=0.8\linewidth]{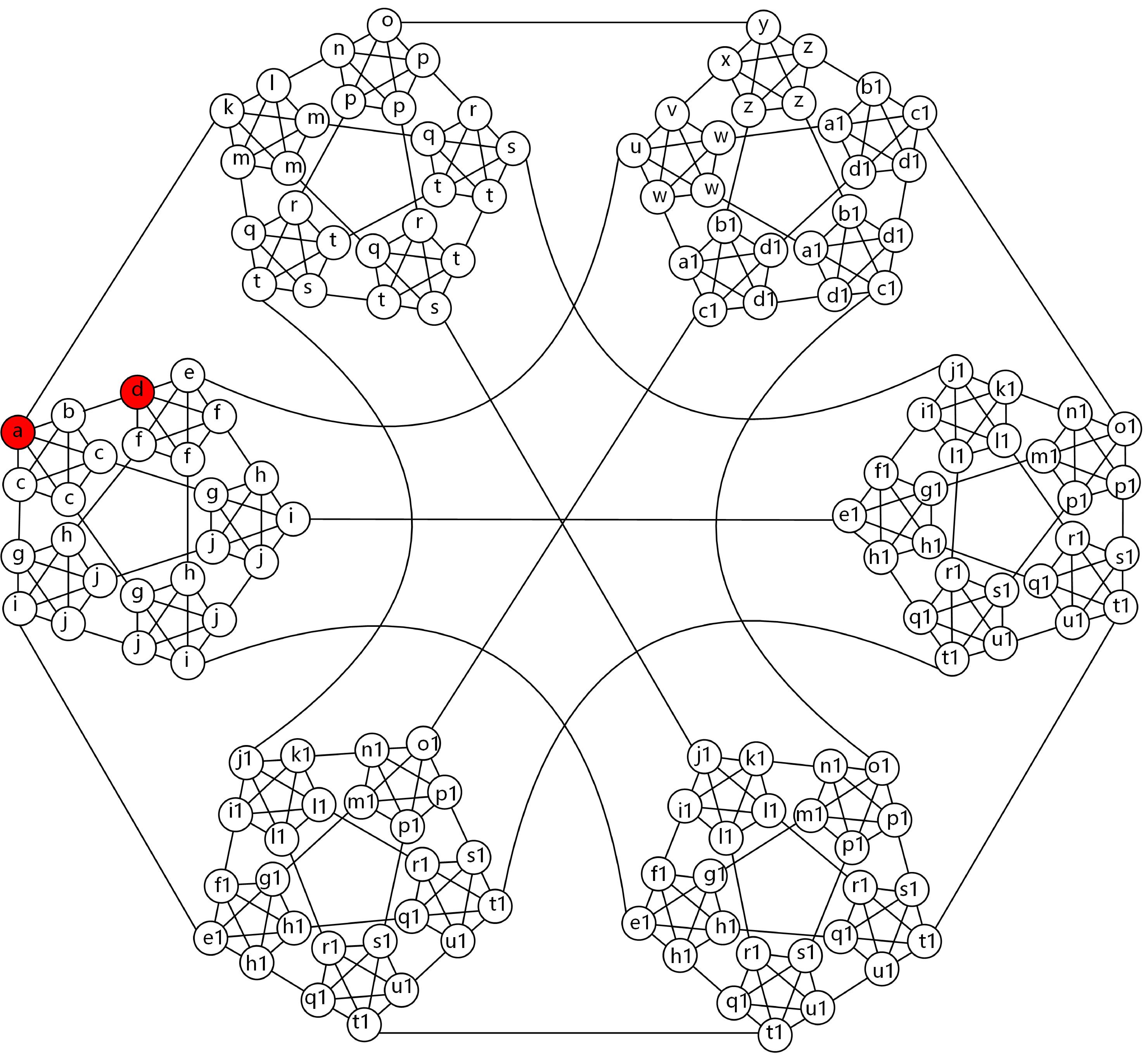}
			\caption{}
			\label{twomarkers2}
		\end{subfigure}
		\begin{subfigure}{0.5\linewidth}
			\centering
			\includegraphics[width=0.8\linewidth]{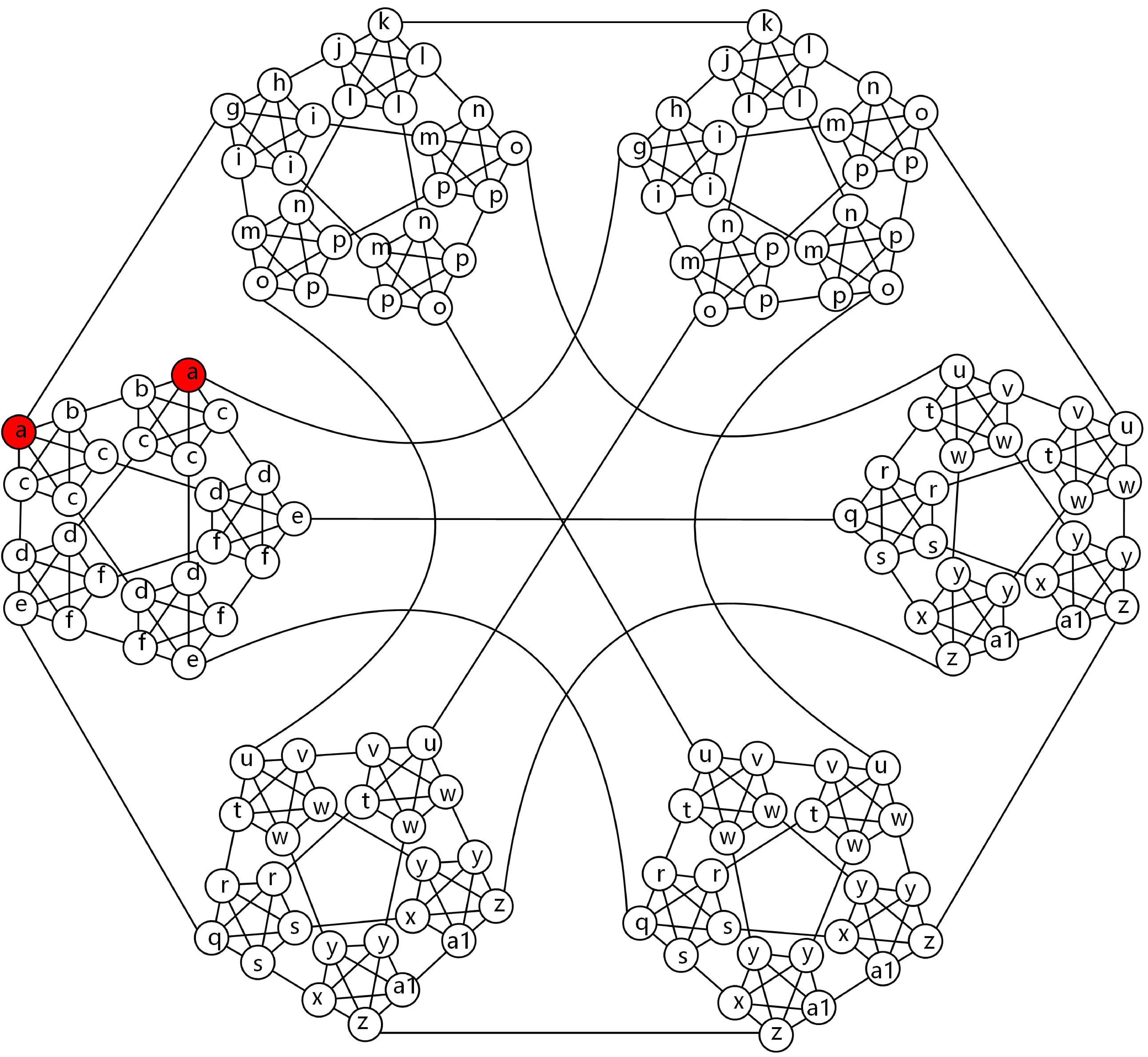}
			\caption{}
			\label{twomarkers3}
		\end{subfigure}
		\begin{subfigure}{0.5\linewidth}
			\centering
			\includegraphics[width=0.8\linewidth]{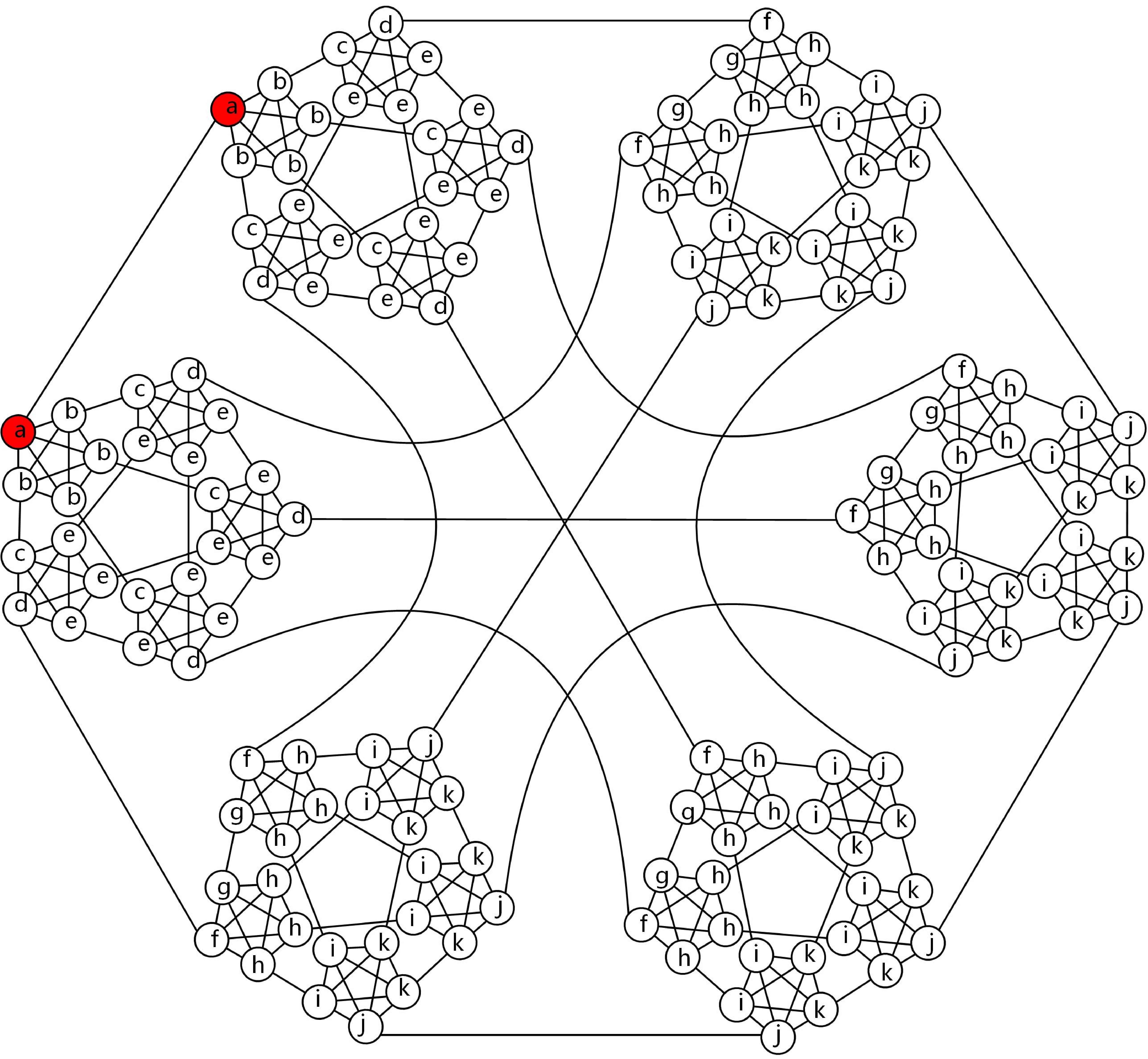}
			\caption{}
			\label{twomarkers4}
		\end{subfigure}
		\begin{subfigure}{1\linewidth}
			\centering
			\includegraphics[width=0.4\linewidth]{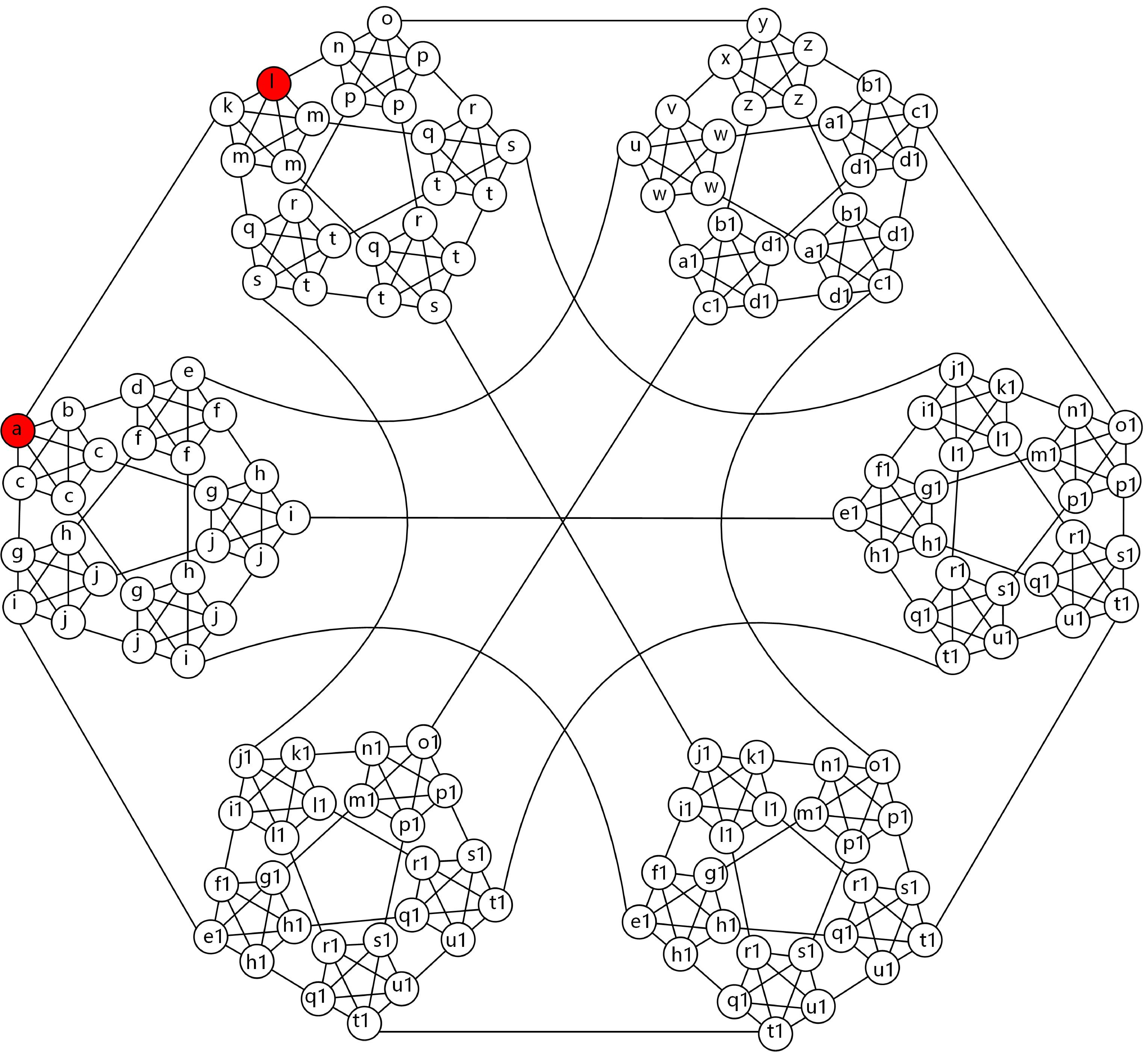}
			\caption{}
			\label{twomarkers5}
		\end{subfigure}
		\caption{Five configurations of the two marked vertices in the second-order lattice, where $M=5$. The vertices that evolve identically are represented by the same letter.}
		\label{twomarkers}
	\end{figure}

	To further investigate the impact of the marked vertices' positions on quantum search, we consider the quantum search on the second-order lattice with two marked vertices. And five different cases are analyzed, as depicted in Fig. \ref{twomarkers}. The corresponding results of the analysis are presented in Table \ref{gammasoftwomarkers}.
	
	\begin{table}
		\caption{The five cases of search with two marked vertices, shown in Fig. $\ref{twomarkers}$, with the subspace dimension, the three stages' critical jumping rate($\gamma_{c}$), and evolution.}\label{gammasoftwomarkers}
		\centering
		\begin{tabular}{p{8em} p{7em} p{10em} p{15em}}
			\hline
			Case & Dimension & $\gamma_c$ & Evolution \\
			\midrule[1pt]
			Figure $\ref{twomarkers1}$ & 47 & $\gamma_{c1,Fig(a)}=5/M$ & $\left| u1 \right\rangle \xrightarrow{} \left| j \right\rangle$ \\
			&  & $\gamma_{c2,Fig(a)}=3/M$ & $\left| j \right\rangle \xrightarrow{} \left| c \right\rangle$ \\
			&  & $\gamma_{c3,Fig(a)}=1/M$ & $\left| c \right\rangle \xrightarrow{} \frac{1}{\sqrt{2}}(\left| a \right\rangle + \left| b \right\rangle)$ \\
			Figure $\ref{twomarkers2}$ & 47 & $\gamma_{c1,Fig(b)}=4/M$ & $\left| u1 \right\rangle \xrightarrow{} \left| j \right\rangle$ \\
			&  & $\gamma_{c2,Fig(b)}=2/M$ & $\left| j \right\rangle \xrightarrow{} \frac{1}{\sqrt{2}}(\left| c \right\rangle + \left| f \right\rangle)$ \\
			&  & $\gamma_{c3,Fig(b)}=1/M$ & $\frac{1}{\sqrt{2}}(\left| c \right\rangle + \left| f \right\rangle) \xrightarrow{} \frac{1}{\sqrt{2}}(\left| a \right\rangle + \left| d \right\rangle)$ \\
			Figure $\ref{twomarkers3}$ & 27 & $\gamma_{c1,Fig(c)}=4/M$ & $\left| a1 \right\rangle \xrightarrow{} \left| f \right\rangle$ \\
			&  & $\gamma_{c2,Fig(c)}=2/M$ & $\left| f \right\rangle \xrightarrow{} \left| c \right\rangle$ \\
			&  & $\gamma_{c3,Fig(c)}=1/M$ & $\left| c \right\rangle \xrightarrow{} \left| a \right\rangle$ \\
			Figure $\ref{twomarkers4}$ & 11 & $\gamma_{c1,Fig(d)}=3/M$ & $\left| k \right\rangle \xrightarrow{} \left| e \right\rangle$ \\
			&  & $\gamma_{c2,Fig(d)}=2/M$ & $\left| e \right\rangle \xrightarrow{} \left| b \right\rangle$ \\
			&  & $\gamma_{c3,Fig(d)}=1/M$ & $\left| b \right\rangle \xrightarrow{} \left| a \right\rangle$ \\
			Figure $\ref{twomarkers5}$ & 47 & $\gamma_{c1,Fig(e)}=3/M$ & $\left| u1 \right\rangle \xrightarrow{} \frac{1}{\sqrt{2}}(\left| j \right\rangle + \left| t \right\rangle)$ \\
			&  & $\gamma_{c2,Fig(e)}=2/M$ & $\frac{1}{\sqrt{2}}(\left| j \right\rangle + \left| t \right\rangle) \xrightarrow{} \frac{1}{\sqrt{2}}(\left| c \right\rangle + \left| m \right\rangle)$ \\
			&  & $\gamma_{c3,Fig(e)}=1/M$ & $\frac{1}{\sqrt{2}}(\left| c \right\rangle + \left| m \right\rangle) \xrightarrow{} \frac{1}{\sqrt{2}}(\left| a \right\rangle + \left| l \right\rangle)$ \\
			\hline
		\end{tabular}
	\end{table}
	
	In Figs. \ref{twomarkers2} and \ref{twomarkers3}, the two marked vertices are located in the same first-order complete subgraph, resulting in $\gamma_{c1,Fig(b)}=\gamma_{c1,Fig(c)}=4/M$ for the first stage. In Figs. \ref{twomarkers4} and \ref{twomarkers5}, the two marked vertices are in two different first-order complete subgraphs, leading to $\gamma_{c1,Fig(d)}=\gamma_{c1,Fig(e)}=3/M$ for the first stage. For the second stage, in Figs. \ref{twomarkers2}-\ref{twomarkers5}, the two marked vertices are situated on different zeroth-order complete subgraphs, yielding $\gamma_{c2,Fig(b)}=\gamma_{c2,Fig(c)}=\gamma_{c2,Fig(d)}=\gamma_{c2,Fig(e)}=2/M$.
	With the above results, we can consider the searches in Figs. \ref{twomarkers2} and \ref{twomarkers3} as equivalent, as well as the searches in Figs. \ref{twomarkers4} and \ref{twomarkers5}.
	Therefore, we conclude that changing the positions of the marked vertices does not affect the search, as long as the number of marked vertices and the corresponding secondary structures remain constant.
	
	In comparison to Figs. \ref{twomarkers2} and \ref{twomarkers3}, the two marked vertices in Fig. \ref{twomarkers1} are positioned on one zeroth-order complete subgraph.
	In the second search stage in Fig. \ref{twomarkers1}, the target secondary structure consists of only one zeroth-order complete subgraph, whereas in Figs. \ref{twomarkers2} and \ref{twomarkers3}, the target secondary structure involves two zeroth-order complete subgraphs. This distinct configuration results in $\gamma_{c2,Fig(a)}=3/M$ in Fig. \ref{twomarkers1}, which is different from $\gamma_{c2,Fig(b)}=\gamma_{c2,Fig(c)}=2/M$ in Figs. \ref{twomarkers2} and \ref{twomarkers3}.
	This local difference has an impact on the overall search process. So in Figs. \ref{twomarkers2} and \ref{twomarkers3}, $\gamma_{c1,Fig(b)}=\gamma_{c1,Fig(c)}=4/M$, while in Fig. \ref{twomarkers1}, $\gamma_{c1,Fig(a)}=(4+1)/M=5/M$. Thus, we conclude that different searches in lower-order substructures will influence the search in higher-order structures.
	
	It is observed that in Figs. \ref{twomarkers4} and \ref{twomarkers5}, the critical jumping rates of each stage are identical to the case with only one marked vertex.
	Furthermore, we have obtained that in Figs. \ref{twomarkers4} and \ref{twomarkers5}, the number of stages, and the time are the same as the ones in the previously discussed cases with one marked vertex.
	This result implies that when the marked vertices are located in different secondary structures, the searches among these substructures does not affect each other.
	In Figs. \ref{twomarkers4} and \ref{twomarkers5}, the first stage of the search can be viewed as two parallel processes searching for a single marked vertex on the second-order lattice. Similarly, in Figs. \ref{twomarkers2} and \ref{twomarkers3}, the two second stages of search processes are equivalent to two parallel processes searching for a single marked vertex on a first-order subgraph.
	
\begin{figure}[htbp]
		\centering
		\begin{subfigure}{0.5\linewidth}
			\centering
			\includegraphics[width=0.6\linewidth]{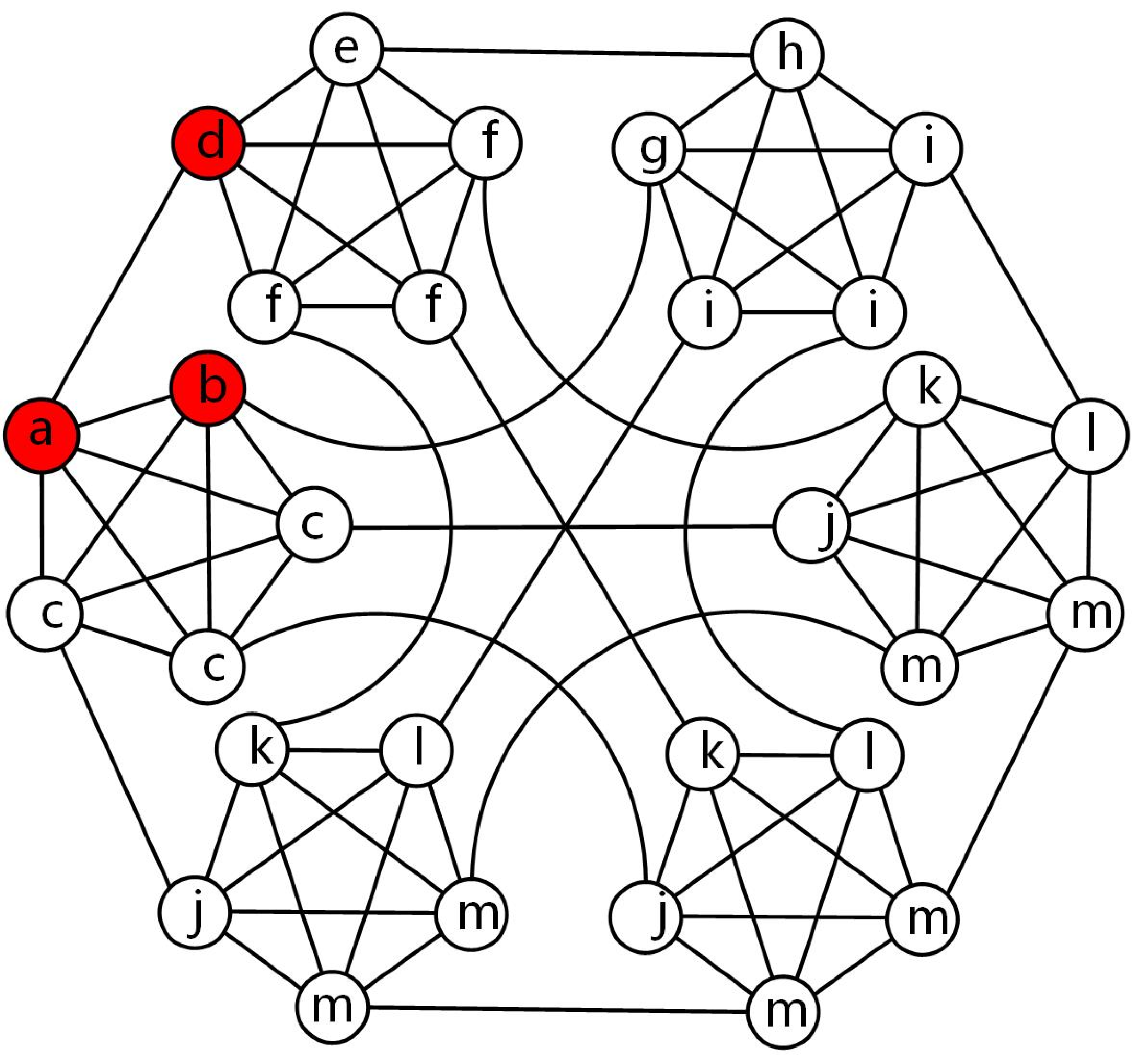}
			\caption{}
			\label{threemarkers}
		\end{subfigure}
		\centering
		\begin{subfigure}{0.4\linewidth}
			\centering
			\includegraphics[width=0.9\linewidth]{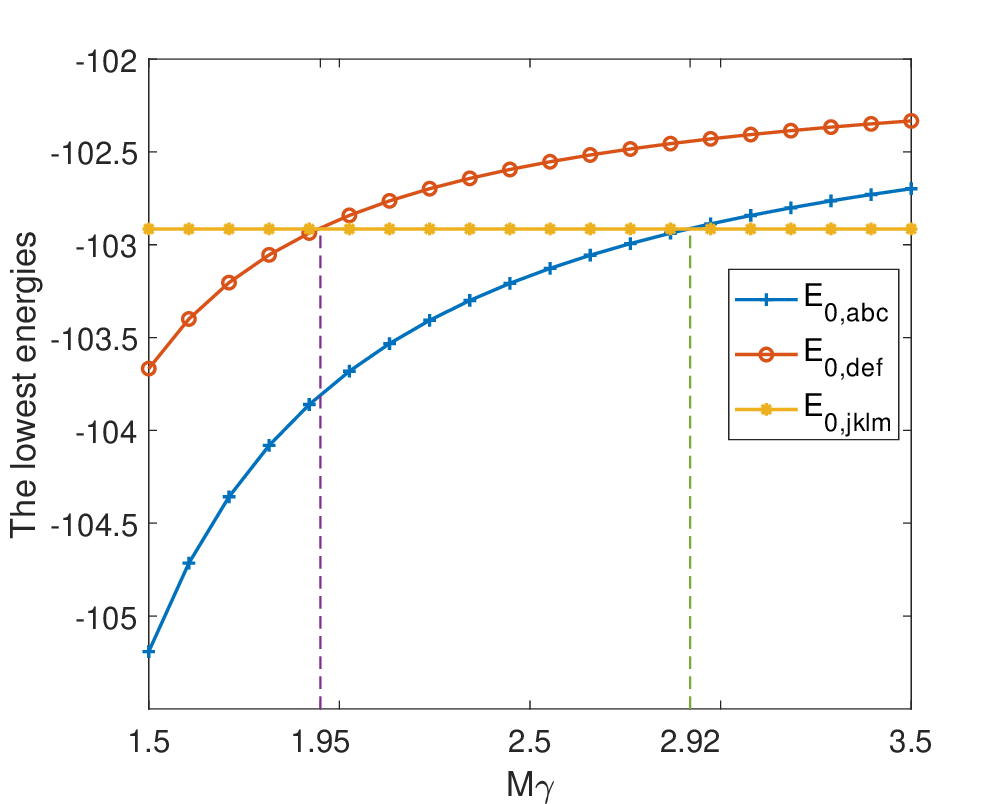}
			\caption{}
			\label{gammasofthreemarkers}
		\end{subfigure}
		\caption{\textbf{a} The scenario of three marked vertices in the first-order lattice. \textbf{b} The critical jumping rates of the first stage.}
		\label{threemarkesfigures}
	\end{figure}

	This observation is further supported by the search illustrated in Fig. \ref{threemarkesfigures}. There are three marked vertices, with $a$ and $b$ situated in the same zeroth-order complete graph, while $d$ is located in another zeroth-order complete graph. When $M$ is large, the initial state $\left| {{\psi}}(0) \right\rangle \approx \left| m \right\rangle$.
	According to the results presented in Fig. \ref{gammasofthreemarkers}, when we want the system to evolve towards $\{\left| a \right\rangle, \left| b \right\rangle\}$, the jumping rate should be $\gamma_{c1,marked(a\&b)} \approx 3/M$. This aligns with the case of having only $a$ and $b$ as marked vertices without the marked vertex $d$ on the first-order lattice.
	When the system evolves towards $\left| d \right\rangle$, the jumping rate $\gamma_{c1,marked(d)} \approx 2/M$. This matches the scenario of having a single marked vertex on the first-order lattice. Therefore, we can conclude that the search for different secondary structures does not interfere with each other when the marked vertices are located in distinct secondary structures.
	
\section{Conclusion}
	\label{sec:conclusion}
	In this paper, degenerate perturbation theory is applied to quantum search on the second-order truncated simplex lattice to determine the critical jumping rate $\gamma_{c}$.  With the $\gamma_{c}$, the system can evolve into the target state at an appropriate time. The construction of the leading-order term of the Hamiltonian must consider the lattice structure. Specifically, when the lattice order exceeds $1$, edges with weight $1$ in the secondary structure, as well as $l$ in $M-l$ and $\sqrt{M-l}$, cannot be omitted.
	
	From the results of the quantum search on the second- and third-order lattices, we have observed that the basis groups can be disregarded if they satisfy the following two constraints:
	(1) They are not part of the shortest path between the initial state and the target state. (2) After the omission of the basis groups, the structural consistency can still be retained.
	By employing this omission, the calculations can be substantially simplified.
	
	We have also shown that the change in the position of the single marked vertex on the second-order lattice has no impact on the three-stage search. To examine the influence of the marked vertices' positions on the search, five distinct configurations of two marked vertices have been studied.
	Our results reveal three rules regarding the impact of different configurations:
	(1) When the number of marked vertices and the secondary structures they are located in remain constant, the variations of their positions do not affect the search. (2) Different searches in lower-order substructures have an influence on the search in higher-order structures. (3) The search for different secondary structures does not interfere with each other when the marked vertices are located on separate secondary structures. Our research provides support for the application of degenerate perturbation theory in continuous-time quantum walk and may be a valuable reference for its implementation in other structures.
	
	\section*{Acknowledgments}
	This work was financially supported by the National Natural Science Foundation of China
	(Grant No. 11965005), and we thank Yunkai Wang for his helpful suggestions.


\end{document}